\documentclass[aps,pra,twocolumn,groupedaddress,floatfix,superscriptaddress]{revtex4-2}

\usepackage[margin=1in]{geometry}
\usepackage[english]{babel}
\usepackage{xcolor}
\usepackage{ucs}
\usepackage{amsmath}
\usepackage{amssymb}
\usepackage{amsthm}
\usepackage{graphicx}
\usepackage{bm}
\usepackage{bbm}
\usepackage{braket}
\usepackage{empheq}
\usepackage{subcaption}
\usepackage{booktabs}
\usepackage{tabularx}
\usepackage[colorlinks=true]{hyperref}
\usepackage[font=footnotesize,labelfont=bf,justification=RaggedRight,singlelinecheck=false]{caption}

\begin{document}

\title{Detecting quantum noise of a solid-state spin ensemble with dispersive measurement}

\author{Mikhail Mamaev}
\thanks{Current affiliation: Department of Physics, University of Toronto. mikhail.mamaev@utoronto.ca}
\affiliation{Pritzker School of Molecular Engineering, University of Chicago, Chicago, Illinois 60637, USA}
\author{Jayameenakshi Venkatraman}
\affiliation{Department of Physics, University of California, Santa Barbara, Santa Barbara, California 93106, USA}
\author{Martin Koppenh\"{o}fer}
\affiliation{Fraunhofer Institute for Applied Solid State Physics IAF, Tullastr.~72, 79108 Freiburg, Germany}
\author{Ania C. Bleszynski Jayich}
\affiliation{Department of Physics, University of California, Santa Barbara, Santa Barbara, California 93106, USA}
\author{Aashish A. Clerk}
\affiliation{Pritzker School of Molecular Engineering, University of Chicago, Chicago, Illinois 60637, USA}

\date{\today}

\begin{abstract}

We theoretically explore protocols for measuring the spin polarization of an ensemble of solid-state spins, with precision at or below the standard quantum limit. Such measurements in the solid-state are challenging, as standard approaches based on optical fluorescence are often limited by poor readout fidelity. Indirect microwave resonator-mediated measurements provide an attractive alternative, though a full analysis of relevant sources of measurement noise is lacking.  In this work we study dispersive readout of an inhomogeneously broadened spin ensemble via coupling to a driven resonator measured via homodyne detection. We derive generic analytic conditions for when the homodyne measurement can be limited by the fundamental spin-projection noise, as opposed to microwave-drive shot noise or resonator phase noise. By studying fluctuations of the measurement record in detail, we also propose an experimental protocol for directly detecting spin squeezing, i.e. a reduction of the spin ensemble's intrinsic projection noise from entanglement.  Our protocol provides a method for benchmarking entangled states for quantum-enhanced metrology.
\end{abstract}

\maketitle

%%%%%
\section{Introduction}
%%%%%

\begin{figure*}
    \center
    \includegraphics[width=0.9\linewidth]{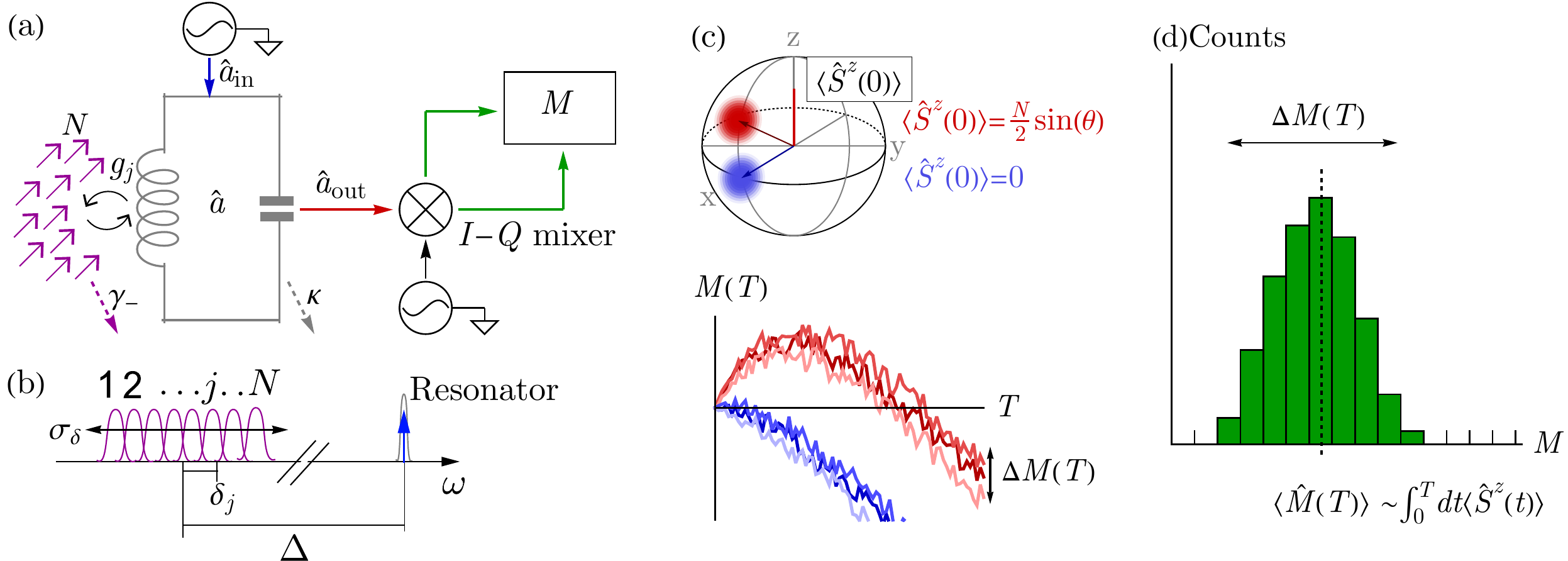}
    \caption{(a) Schematic of a circuit-QED setup relevant to our system, with $N$ spin-1/2 degrees of freedom coupling to a resonator $\hat{a}$ with individual couplings $g_j$. Excitations decay from the resonator to its environment with strength $\kappa$, and from individual spins with rate $\gamma_{-}$. The resonator is driven by an input microwave field $\hat{a}_{\mathrm{in}}$. The quadratures of the light field coming out of the resonator $\hat{a}_{\mathrm{out}}$ are passed through, e.g., an $I$-$Q$ mixer, and measured to obtain a time-integrated homodyne current $M$. (b) Schematic of the system spectrum. The spins have inhomogeneous frequencies $\delta_j$, while the resonator is detuned by $\Delta$ from the spins' center frequency. (c) Schematic of how an initial collective spin polarization $\langle \hat{S}^{z}(0)\rangle$ (either on the equator of the spin Bloch sphere, or inclined by some angle $\theta$) affects the dynamics of the integrated homodyne quadrature measurement outcome $M(T)$ over a collection time $T$. Different trajectories correspond to different runs of an experiment. (d) For a fixed set of $N_{\mathrm{runs}}$ experimental runs, the distribution of measurement outcomes after a collection time $T$ may be binned. The mean of the distribution is the expectation value $\langle \hat{M}(T)\rangle$ dependent on the collective spin polarization. The width of the distribution is set by the variance $(\Delta M(T))^2$, which we seek to make spin-projection-noise limited.}
    \label{fig:fig1}
\end{figure*}

Surpassing the standard quantum limit (SQL) of measurement sensitivity is an important frontier of quantum metrology~\cite{caves1981,clerk2010,aasi2013}. 
For matter-based sensors, the SQL is the fundamental physical sensitivity limit if only uncorrelated or classical states are used~\cite{pezze2018}. 
Provided technical readout noise is below the SQL, generating entanglement in an ensemble of atom-based or spin-based sensors can enable enhanced precision, e.g., by squeezing the quantum projection noise of the sensing state~\cite{kitagawa1993seminal,pezze2018,ma2011squeezingReview}. 
Pioneering experiments have demonstrated squeezing in atomic ensembles~\cite{leroux2010,Riedel2010,Gross2010} and in NMR experiments~\cite{auccaise2015spinSqueezingSS}, and have already created new standards for timekeeping~\cite{eckner2023}. 
Solid-state spin sensors, such as the NV center in diamond, form a prime testbed of these phenomena and they provide the unique opportunity to sense a wide range of condensed-matter~\cite{jenkins2022imaging}, geological~\cite{glenn2017micrometer}, and biological~\cite{aslam2023quantum} quantities with sensitivity beyond the SQL.
Several theoretical proposals have been made to generate metrologically useful entanglement in solid-state spin ensembles~\cite{bennett2013phonon,DallaTorre2013,borregaard2017,groszkowski2020,groszkowski2022,zheng2022,Block2024}, and a recent experiment demonstrated spin-squeezing in an ensemble of dipolar interacting NV centers, confirmed via an indirect detection method~\cite{wu2025squeezingEnsemble}.

However, leveraging spin squeezing for actual sensitivity improvements remains hindered by the fact that measurements are typically limited by technical noise (such as photon shot noise in standard NV readout by optical fluorescence) rather than by the fundamental quantum projection noise~\cite{barry2020review}. 
Hence, there is a strong motivation to realize spin-projection-noise-limited readout of spin ensembles or at least to mitigate detection noise~\cite{koppenhofer2022dissipative,koppenhoefer2023}. 
A recent heroic effort~\cite{maier2025} measured an NV ensemble ($N\sim 30$) at the quantum projection noise limit by leveraging a nuclear-spin-assisted repetitive readout scheme~\cite{jiang2009}. However the experiment necessitated high (2.7 T) magnetic fields and high (80 GHz) microwave signals, which -- alongside scaling to larger $N$ -- is technically challenging.

A promising alternative is indirect readout via transduction of the spin polarization into an auxiliary mode~\cite{koppenhoefer2023omit}, such as a resonator, and the achievable noise limits have also been explored for single-spin systems~\cite{danjou2019dispersiveSingle, yang2024exceptional}.
There have been pioneering experiments detecting spins in a variety of solid-state hosts using resonator-assisted readout~\cite{bienfait2016controlling,eisenach2021cavity,albertinale2021detecting, wen2025addressing}, including direct measurement of the collective polarization of a spin ensemble~\cite{ebel2021dispersive} and even single spin detection~\cite{wang2023}. 
Unfortunately, a measurement precision saturating the SQL of a solid-state spin ensemble with resonator-assisted readout is yet to be demonstrated. 
Part of the challenge is that, even on a theoretical level, the system exhibits a large and complex parameter space, rendering it difficult to find an experimentally optimal regime to operate in. 
There exist parametric conditions quantifying the achievable sensitivity with resonator-assisted readout~\cite{barry2024sensitive}, but they generally invoke empirical scaling factors to characterize the measurement efficiency. 
An outstanding set of questions in the field is thus: i) What are the parametric requirements on an ensemble and resonator to be able to reliably detect spin projection noise, and ii) How can such information be used to characterize the presence of entanglement such as squeezing?

In this work, we answer these questions via a theoretical study of a dispersive readout scheme~\cite{blais2004reviewDispersive} of the collective spin polarization of a solid-state spin ensemble coupled to a microwave resonator. Specifically, we consider using a dispersive shift $\sim \chi\hat{a}^{\dagger}\hat{a} \hat{S}^{z}$ (with resonator mode $\hat{a}$ and dispersive spin-resonator coupling $\chi$) to extract information on the collective spin component $\hat{S}^{z}$ by driving $\hat{a}$, and measuring the change in reflected output due to $\hat{S}^{z}$ via homodyne readout.
Dispersive readout using microwave resonators is a well-established technique for circuit-QED platforms~\cite{blais2004reviewDispersive}, where almost ideal measurements of one or a few qubits are straightforward to realize. The situation is quite different for solid-state many-spin sensors.  Here, one has to contend with disorder (inhomogeneous broadening), something which is ubiquitous in solid-state systems~\cite{stanwix2010coherence, krimer2014nonMarkovianBroadening}.  Further, the experimentally attainable dispersive shifts are dramatically smaller due to small spin-resonator magnetic dipole-mediated couplings.

Due to the longer measurement timescales involved, the effects of intrinsic $T_1$ decay of the spins must be accounted for, as they dynamically change both the measured signal and the noise contributions mid-measurement. Our analysis considers how $T_1$ decay of a large spin ensemble changes the fluctuation properties of the ensemble as it evolves during the measurement, and also accounts for the effects of inhomogeneous broadening. The analysis also includes both the effects of photon shot noise in the driven resonator, as well as resonator phase noise.  We derive a general analytic condition for when the measurement can be made spin-projection-noise limited.
We also carefully analyze the requirements for being able to detect spin squeezing in the initial state of the spin ensemble via a homodyne measurement, deriving necessary parametric conditions on the setup that need to be satisfied.

%%%%%
\section{Theoretical model}
\label{sec:theoretical-model}
%%%%%

We consider a generic spin ensemble plus resonator setup, comprised of $N$ spin-1/2 degrees of freedom coupled to a single resonator, as depicted in Fig.~\ref{fig:fig1}(a).
We consider the system to exhibit inhomogeneous broadening, i.e., the transition frequencies of individual spins are spread stochastically across a frequency range rather than being uniform. The physics of the system can be modeled by the following Hamiltonian and quantum master equation in a frame rotating at the mean spin-transition frequency,
\begin{equation}
\label{eq:1}
\begin{aligned}
\hat{H} &= \Delta \hat{a}^{\dagger}\hat{a} + \sum_{j=1}^N \delta_j \hat{s}_j^{z} + \sum_{j=1}^N \left(g_j \hat{a}^{\dagger}\hat{s}_{j}^{-} + \mathrm{h.c.}\right),\\
\frac{d}{dt}\rho &= - i [\hat{H},\rho] + \kappa \mathcal{D}[\hat{a}]\rho + \gamma_{-} \sum_{j=1}^N \mathcal{D}[\hat{s}_{j}^{-}]\rho.
\end{aligned}
\end{equation}
Here, $\hat{a}$ is the lowering operator for the resonator mode and $\hat{s}_j^{+}$, $\hat{s}_j^{-}$, $\hat{s}_j^{z}$ are spin-1/2 operators for spin $j$ with commutation relations $[\hat{s}_j^{+},\hat{s}_j^{-}] = 2 \hat{s}_j^{z}$. As depicted in Fig.~\ref{fig:fig1}(b), the resonator has a detuning $\Delta$ from the spin transition center frequency. The spins themselves have random detunings $\delta_j$ from this center frequency, drawn from a generic probability distribution with characteristic width $\sigma_{\delta}$ and mean zero (such as a Gaussian with variance $\sigma_{\delta}^2$). In this parametrization, $\sum_j \delta_j = 0$. Each spin couples to the resonator with strength $g_j$. We also consider the spins to have an intrinsic energy relaxation rate $\gamma_{-}$, and the resonator has a photon loss rate $\kappa$. In Table~\ref{table_Params}, we show all relevant parameters for the system we study, along with realistic values for a sample experimental implementation employing NV centers in a diamond substrate coupled to a superconducting microwave resonator patterned on top of it. The details of this candidate platform are described in the next section. These are the values used in all subsequent plots and simulations unless otherwise specified.

\begin{table*}[]
\caption{System parameters. All energies are in units of $s^{-1}$, setting $\hbar = 1$ (can be converted to Hz by removing the factors of $2\pi$). All numerical calculations and plots use these parameters, unless otherwise specified.}
\begin{tabular}{l|l|l}
\toprule
Symbol & Meaning & Value \\
\midrule
$N$ & Atom number & $10^6$ \\ 
$g_j$ & Spin-resonator coupling & $2\pi \times 50$ $s^{-1}$ \\ 
$\sigma_{\delta}$ & Inhomogeneous broadening width & $2\pi \times 10^6$ $s^{-1}$ \\ 
$\delta_j$ & Spin frequency & $\in[-\sigma_{\delta},\sigma_{\delta}]$ \\ %\hline
$\gamma_{-}$ & Spin decay rate = $T_1^{-1}$ & $2\pi \times 1$ $\mathrm{s}^{-1}$ \\ %\hline
$\kappa$ & Resonator loss rate & $2\pi \times 10^5$ $s^{-1}$ \\ %\hline
$\Delta$ & Resonator-spin detuning & $2\pi \times 5\times 10^6$ $s^{-1}$ \\ %\hline
$\overline{n}$ & Resonator mean photon number $=\langle\hat{a}^{\dagger}\hat{a}\rangle$ & $10^5$ \\ %\hline
$\gamma_j$& Renormalized spin decay rate $= \gamma_{-} + \kappa g_j^2/(\Delta-\delta_j)^2$ & $2\pi \times 1$ $\mathrm{s}^{-1}$ \\ %\hline
$\chi_j$ &  Spin dispersive coupling $=g_j^2/(\Delta-\delta_j)$ & $2\pi \times 5 \times 10^{-4}$ $s^{-1}$ \\ %\hline
$T$ & Homodyne measurement time & Varies \\
$\lambda$ & Parameter characterizing dispersive measurement & $16 \chi^2 \overline{n} N / (\kappa \gamma)$  \\
&  quality  &    \>\>\>(for homogeneous $\chi_j = \chi$, $\gamma_j = \gamma$ )\\
\bottomrule
\end{tabular}
\label{table_Params}
\end{table*}

We focus on the dispersive regime, which assumes that the spin-resonator detuning is much larger than the spin-resonator coupling.
Compared to resonantly-coupled spins, this approach trades coupling strength for improved control over signal dynamics, and benefits from access to the long spin relaxation times attainable by, e.g., cryogenic NV centers. Furthermore, due to inhomogeneous broadening of spin frequencies, most of a spin ensemble can be dispersive even if one tries to drive the system resonantly; we will discuss such effects further on.
Specifically, as detailed in the next section, for readout purposes we will apply a drive to the resonator to create a nonzero average cavity photon number $\overline{n} = \langle\hat{a}^{\dagger}\hat{a}\rangle$; the condition for being in the dispersive limit is then $|\Delta-\delta_j| \gg \sqrt{\overline{n}}|g_{j}|$ for all $j$. In this regime, direct transfer of excitations from the spins to the resonator is energetically suppressed, allowing us to neglect many of the more complicated dynamics that can be induced by the resonator. The leading-order perturbative effect of the coupling is a dispersive interaction, which can be written as (see Appendix~\ref{app_SchriefferWolff}),
\begin{equation}
\begin{aligned}
\label{eq_SWModel}
\hat{H}_{\mathrm{SW}} &=\Delta \hat{a}^{\dagger}\hat{a} + \sum_{j=1}^N \delta_j \hat{s}_{j}^{z} - 2\hat{a}^{\dagger}\hat{a} \sum_{j=1}^N \chi_j \hat{s}_{j}^{z},\\
\frac{d}{dt}\rho &= - i [\hat{H}_{\mathrm{SW}},\rho] + \kappa \mathcal{D}[\hat{a}]\rho + \sum_{j=1}^N \gamma_j \mathcal{D}[\hat{s}_{j}^{-}]\rho,\\
\end{aligned}
\end{equation}
where the coefficients are,
\begin{align}
    \chi_j &= \frac{|g_j|^2}{\Delta -\delta_j}, &
    \gamma_j &= \gamma_{-} + \frac{\kappa |g_j|^2}{(\Delta - \delta_j)^2}.
\label{definition_chij_gammaj}
\end{align}
The dispersive shift stems from the coupling term $\sim \hat{a}^{\dagger}\hat{a} \sum_{j} \chi_j\hat{s}_j^z$. Crucially, in deriving the above model, we make the experimentally-relevant assumption that the inhomogeneous broadening width $\sigma_{\delta}$ is \textit{strong} compared to all of the resonator-mediated processes. Formally, we make the approximation that $|g_j^{*} g_{j'}|/|\Delta - \delta_j| \ll |\delta_j - \delta_{j'}| \approx \sigma_\delta$ for all $j \neq j'$, which says that the spins are far apart in frequency (due to the inhomogeneous broadening) compared to their resonator-mediated coupling, and hence cannot undergo flip-flop interactions or exhibit superradiant enhancement of Purcell decay (Appendix~\ref{app_SchriefferWolff} provides details on this approximation). There is however a non-collective Purcell decay of each spin, which will play a role in the detection properties of the system.

\begin{figure}
    \centering
    \includegraphics[width=\linewidth]{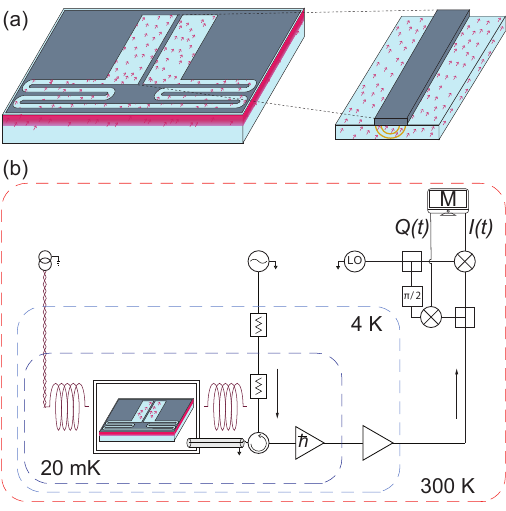}
    \caption{Experimental setup for spin ensemble detection at and beyond the standard quantum limit (SQL): (a) proposed experimental device and (b) measurement setup. (a) To realize Fig. \ref{fig:fig1} (a), a superconducting thin film (grey) is lithographically patterned on top of the diamond substrate (light blue) containing a near-surface defect center spin ensemble (pink). The patterned superconductor defines a lumped-element resonator circuit with a narrow wire forming the inductor (L) carrying current, and planar meandered fingers forming the capacitor (C) carrying charge, with the resonant frequency lying in the microwave regime. A zoomed-in cartoon around the wire is shown as inset. The current-carrying wire produces a microwave magnetic field (gold lines) which couples to the spins located within the microwave resonator mode volume. This coupling is modeled by the Jaynes-Cummings interaction written in Eq. \ref{eq:1} and ultimately responsible for the dispersive interaction in Eq. \ref{eq_SWModel}. b) Microwave detection circuit of the device shown in (a) and caricatured in Fig. \ref{fig:fig1}. The Larmor precession frequency of the spins is tuned by a DC magnetic field produced by a Helmholtz coil (maroon). The measurement chain is kept low-noise by attenuating thermal noise along the input lines and adding low-noise amplification along the output lines. Homodyne microwave detection measures the field quadratures $I$ and $Q$ by interfering the signal exiting the resonator with a known local oscillator (LO) reference.  Time-integration of the field quadratures yields the measurement observable $M$.}
    \label{fig:sample}
\end{figure}

There can also be flip-flop interactions $\sim \hat{s}_j^{+}\hat{s}_{j'}^{-}$ originating from direct dipole-dipole couplings between the spins.  If the ensemble is sufficiently spatially dense, these couplings are not necessarily slow compared to the timescales for measurement. For instance, dipole-dipole coupling strengths of NV centers in diamond are $\sim 50 \,\mathrm{MHz} \times (\text{nm})^3$~\cite{hall2016detection, zu2021emergent}, or $\sim 50$ kHz for vacancies $10$ nm apart. These interactions are generally viewed as a source of decoherence, though there are recent explorations of their effects on superradiant dynamics~\cite{kersten2026self} and on using them for entanglement generation~\cite{wu2025squeezingEnsemble}. Such dipole-dipole couplings do not affect our subsequent results because they will not appreciably change the quantum dynamics of the total spin polarization observable $\sum_j \chi_j \hat{s}_j^z$ we seek to measure. Specifically, such flip-flop interactions are only relevant for spins $j$ and $j'$ that have similar frequencies $\delta_j \sim \delta_{j'}$ (compared to the flip-flop strength), which is unlikely for ensembles with $\sigma_{\delta} \sim $ MHz inhomogeneous broadening. Even if flip-flops do happen, their overall dispersive shift $\sim \chi_j\hat{s}_j^z + \chi_{j'}\hat{s}_{j'}^z$ will be unaffected provided $\chi_j$ and $\chi_{j'}$ are comparable, since the total spin polarization $\hat{s}_j^z+\hat{s}_{j'}^z$ commutes with the flip-flop interaction $\sim (\hat{s}_j^{+}\hat{s}_{j'}^{-}+\mathrm{h.c.})$, and does not experience any additional dynamics induced by the dipole-dipole interactions.

Note that if the couplings $\chi_j$ \textit{are} inhomogeneous while spin frequencies are sufficiently close, dipole-dipole interactions \textit{will} affect the signal we are able to measure (since it is a weighted sum of contributions from different spins, rather than a collective spin polarization). The parameter regime of strong inhomogeneity will exhibit a reduced signal-to-noise ratio for standard, Ramsey-style interferometric measurement protocols; such effects will be explored in Section~\ref{subsec_Inhomogeneity}.

\section{Model experimental system}
\label{sec_expt}
To realize the model spin-resonator system presented in Section~\ref{sec:theoretical-model} and caricatured in Fig.~\ref{fig:fig1}, we introduce a concrete example shown in Fig.~\ref{fig:sample}. The system consists of a superconducting  microwave resonator inductively coupled to a spin ensemble, which, for the sake of concreteness, we take to be an ensemble of NV centers in diamond. 
The microwave resonator consists of a planar inductor-capacitor (LC) circuit with a wire constituting the inductor and meandered fingers defining the planar capacitor (C). The flow of current through the wire and the oscillation of charge between the planar capacitor plates define the resonance frequency of the circuit as $\omega_r = 1/\sqrt{LC}$.
A DC magnetic field controls the detuning $\Delta$ between the spins and resonator.
Spins near the wire inductively drive current in the wire much like in nuclear-magnetic resonance (NMR) or electron paramagnetic resonance (EPR).
The task of achieving spin projection noise-limited sensitivity is translated to quantum-limited detection of the microwave field.

To provide a scale for the coupling between the resonator and a single-spin, we note that, when the microwave resonator is in its ground state (which occurs for milliKelvin operating temperatures for $\omega_r/2\pi \sim \mathcal{O}(\mathrm{GHz})$), the current flowing through the wire is given by the zero-point current $ i_{\mathrm{zpf}} = \omega_r \sqrt{\hbar / 2 Z_0} \approx 50~\mathrm{nA}$ for a resonator at frequency $\omega_r/2\pi \approx 5~\mathrm{GHz}$ with impedance $Z_0 \approx 20~\mathrm{\Omega}$. This current is associated with a zero-point magnetic field given by the Biot-Savart law as
$B_{\mathrm{zpf}} = \mu_0 i_{\mathrm{zpf}}  / (2 \pi r)$ at a distance $r$ from the wire in the far-field limit, where $\mu_0$ is the permeability of free space. For $B_{\mathrm{zpf}}$ oriented entirely perpendicular to a resonant single electron spin, the resulting spin-photon coupling would be given by $g/2\pi = B_{\mathrm{zpf}} \langle 0 | \gamma_e \sigma_x |1 \rangle \approx 50~\mathrm{Hz}$. 

Fig~\ref{fig:sample}(b) shows a quantum-limited detection chain to detect the spin-driven current with the LC circuit thus expanding upon the cartoon schematic in Fig.~\ref{fig:fig1}(a). By submitting the resonator to a microwave tone and measuring the phase of the output microwave irradiation relative to the reference input through a homodyne measurement scheme provides access to the collective spin polarization in the dispersive regime.

In the next section we model the readout chain in detail and derive the criterion for resolving the quantum projection noise above technical noise introduced by the measurement setup.

%%%%%
\section{Homodyne detection and observation of spin-projection noise}
\label{sec:detection}
%%%%%

In a standard Ramsey-style sensing protocol, one seeks to measure some unknown phase $\theta$ caused by, e.g., precession of spins in a magnetic field. A Ramsey sequence translates this phase information into the ensemble's collective spin polarization $\sum_j \hat{s}_j^z$, which can then be read out.
We will do this using the resonator and the overall dispersive coupling $-2\hat{a}^{\dagger}\hat{a}\sum_j\chi_j \hat{s}_j^z$.
As per standard dispersive measurement protocols, the dispersive coupling leads to a spin-dependent change of the effective frequency of the resonator. This change can be detected by measuring the reflection phase of a microwave tone driving the resonator. Concretely, the change is measured via an effective homodyne detection, i.e., measuring one quadrature of the outgoing microwave field (with an $I-Q$ mixer one can also perform heterodyne measurements of both quadratures, although this is not necessary for our goal of just measuring the spin polarization) (see Fig. \ref{fig:fig1}). While this approach is well established in the field, we will focus on explicitly characterizing how the intrinsic quantum noise of the spin ensemble's initial state dynamically manifests itself in the output-field measurement. We will also consider the interplay of the quantum noise with intrinsic $T_1$ decay of the spins happening in parallel to the measurement, and determine the optimal timescales and parameters needed to reach spin-projection-noise-limited precision.

%%%
\subsection{Integrated homodyne quadrature}
%%%

We assume that the resonator is driven by a resonant monochromatic tone (same frequency as the resonator in the absence of spins), described by a bosonic input field $\hat{a}_{\mathrm{in}}$, with an effective drive Hamiltonian $-i\sqrt{\kappa_c}\left(\hat{a}^{\dagger}\hat{a}_{\mathrm{in}}-\mathrm{h.c.}\right)$, where $\kappa_c$ is the coupling between the resonator and driving tone. The Heisenberg equation of motion for the resonator field $\hat{a}$, including both the Hamiltonian and dissipator from Eq.~\eqref{eq_SWModel} and this incident driving tone, is given by
\begin{equation}
\frac{d}{dt}\hat{a}(t) = \left(- \frac{\kappa}{2} + 2i \sum_{j=1}^N \chi_j \hat{s}_j^z\right)\hat{a}(t) - \sqrt{\kappa_c}\hat{a}_{\mathrm{in}}(t).
\end{equation}
The reflected light field coming back from the resonator, described by bosonic output field $\hat{a}_{\mathrm{out}}$, can be written via standard input-output theory as,
\begin{equation}
\hat{a}_{\mathrm{out}}(t) = \hat{a}_{\mathrm{in}}(t) + \sqrt{\kappa_c}\hat{a}(t).
\end{equation}
We assume that loss from the resonator is dominated by the coupling to the measured channel (i.e., the resonator is overcoupled) and set $\kappa_c = \kappa$ going forward, although our results do not qualitatively change if this assumption does not hold, beyond re-scaling the ensuing signal by a factor $\kappa_c / \kappa$ with $\kappa_c$ the loss into the measured channel only. We seek to measure the quadrature of this output field sensitive to the spin polarization, which is (based on choice of drive phase without loss of generality, see Appendix~\ref{app_InputOutputHomodyne}):
\begin{equation}
\hat{I}_{\mathrm{out}}(t) = \hat{a}_{\mathrm{out}}(t) + \hat{a}_{\mathrm{out}}^{\dagger}(t).
\end{equation}
The goal is to extract as much information as possible about the initial state of the spins from the dynamics of this quadrature. The simplest approach is a time-integrated homodyne protocol, for which we integrate the quadrature over some collection time $T$. The time-integrated observable is:
\begin{equation}
\begin{aligned}
\hat{M}(T) &= \int_0^T dt \>\hat{I}_{\mathrm{out}}(t).
\end{aligned}
\end{equation}
The expectation value of this observable has units of $\sqrt{\mathrm{time}}$ (reflecting the fact that photon shot noise will cause its variance to grow diffusively in time). There also exist more sophisticated protocols that use a filter for the time-integration to pick out intervals of time where the strongest signal manifests~\cite{Gambetta-PhysRevA.76.012325}, although we will not consider such techniques in this work.

We operate in the regime where the resonator dynamics are much faster than any spin exchange, $\kappa \gg \chi_j,\gamma_j$, and the resonator quickly builds up to a driven-dissipative equilibrium photon number $\overline{n} = \langle\hat{a}^{\dagger}\hat{a}\rangle$, at which it remains fixed throughout any dynamics we study. Under these assumptions, the time-integrated quadrature expectation value is proportional to the spin polarization (see Appendix~\ref{app_InputOutputHomodyne} for details),
\begin{equation}
\label{eq_Signal}
\langle\hat{M}(T)\rangle= \frac{8\sqrt{\overline{n}}}{\sqrt{\kappa}}\sum_{j=1}^N \chi_j \int_0^T dt \>\langle\hat{s}_j^z (t)\rangle.
\end{equation}
Under the dispersive model, the dynamics of the spin-polarization observable $\langle \hat{s}_j^z(t)\rangle$ are set only by the $T_1$ decay and single-spin Purcell decay, which lead to a combined decay rate $\gamma_j$ defined in Eq.~\eqref{definition_chij_gammaj}. The resulting time-evolution can be written explicitly:
\begin{align}
\langle \hat{s}_j^z(t)\rangle &= -\frac{1}{2}  + \frac{e^{-\gamma_j t}}{2}\left(1 + 2 \langle\hat{s}_{j}^{z}(0)\rangle\right),\\
\int_0^{T}dt \langle\hat{s}_j^{z}(t)\rangle &=\frac{1}{\gamma_{j}}\left(\langle\hat{s}_j^{z}(0)\rangle + \frac{1}{2}\right)\left(1-e^{-\gamma_j T}\right) - \frac{T}{2},
    \label{int_expectation_szt}
\end{align}
with $\langle\hat{s}_{j}^z(0)\rangle$ the initial spin polarization. By measuring $\langle\hat{M}(T)\rangle$ and comparing it to Eq.~\eqref{int_expectation_szt}, we gain information about the initial condition of the spins.

More specifically, a single run of an experiment measuring the integrated homodyne current for some initial condition of the spins will report a single value $M(T)$ at each time $T$. For a set of $N_{\mathrm{runs}}$ experimental runs, as depicted in Fig.~\ref{fig:fig1}(c), we will have a set of values $\{M(T)\}$ with some stochastic distribution due to noise. As depicted in Fig.~\ref{fig:fig1}(d), the distribution may be binned. The mean of the distribution is the expectation $\langle\hat{M}(T)\rangle$, which depends on (and thus provides a measurement of) the spin initial condition. The variance of the distribution will be given by,
\begin{equation}
(\Delta M(T))^2 = \langle \hat{M}(T)^2\rangle - \langle \hat{M}(T)\rangle^2.
\end{equation}
The variance characterizes the total noise affecting the measurement, including both spin projection noise and other factors. A first experiment seeking to verify that the measurement is spin-projection-noise limited may simply measure $(\Delta M(T))^2$ by computing the width of the resulting histogram for various $T$, and seeing if it changes in the presence of spins. In this context, the measurement variance itself is akin to an observable that can be measured in its own right. Should the measurement indeed be spin-projection noise limited, we can also explore further reductions due to entanglement such as spin-squeezing; this will be explored in Section~\ref{sec:Entanglement}.

It is also important to note that what we will end up measuring  with the dispersive approach is not necessarily the fully-collective initial spin polarization $\sum_j \langle \hat{s}_j^z(0)\rangle$, but rather a sum weighted by the dispersive coupling strengths of the spins $\sum_j\chi_j \langle\hat{s}_j^{z}(0)\rangle$. The observable we will have access to is the collective spin polarization plus a correction term due to disorder in $\chi_j$, which will eventually limit the achievable measurement precision; the effects of such inhomogeneity will be discussed further in Section~\ref{subsec_Inhomogeneity}.

\begin{figure*}
    \centering
    \includegraphics[width=0.7\linewidth]{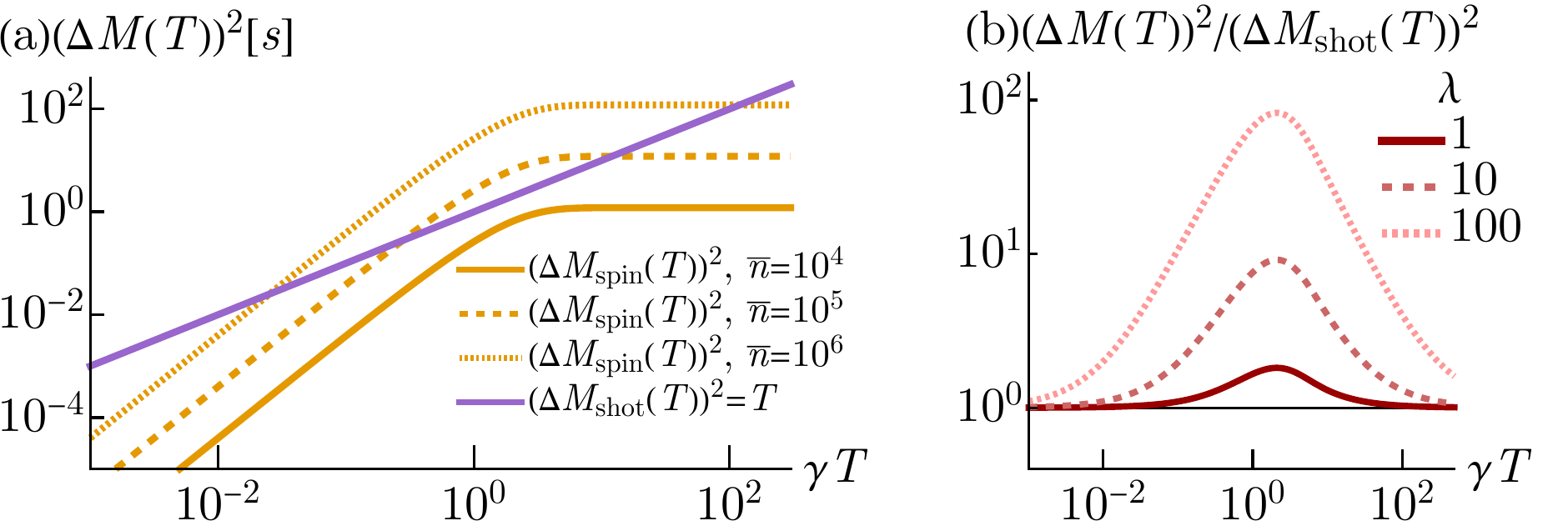}
    \caption{(a) Contributions to the measurement variance $(\Delta M(T))^2 = (\Delta M_{\mathrm{spin}}(T))^2+(\Delta M_{\mathrm{shot}}(T))^2$ [defined in Eq.~\eqref{eq_NoiseTotal}] from intrinsic spin-projection noise $(\Delta M_{\mathrm{spin}}(T))^2$ and shot noise $(\Delta M_{\mathrm{shot}}(T))^2 = T$ from homodyne detection. We use parameters from Table~\ref{table_Params}, but vary the number of resonator photons $\overline{n}$. (b) Total measurement variance $(\Delta M(T))^2$, for different values of the measurement-quality parameter $\lambda$, defined in Eq.~\eqref{eq_FinalSmallParameter}, normalized by the shot noise. We use the same parameters as panel (a), varying $\lambda$ by tuning $\overline{n}=2.5 \times 10^4, 10^5, 10^6$ for $\lambda =1, 10,100$.}
    \label{fig:SignalAndNoise}
\end{figure*}

%%%%%
\subsection{Variance of homodyne measurement}
%%%%%

Before addressing a specific signal to measure (e.g., a phase imprinted on the spins), we first consider the quantum noise of the measurement itself. We want to understand what the requirements are to have the variance $(\Delta M(T))^2$ be dominated by spin-projection noise, rather than shot noise in the homodyne current.
We assume that the drive tone on the resonator is in a coherent state $\hat{a}_{\mathrm{in}}(t) = \alpha_{\mathrm{in}} + \hat{\xi}(t)$ with coherent amplitude $\alpha_{\mathrm{in}}$ and intrinsic quantum fluctuations $\hat{\xi}$. The latter are assumed to have white-noise correlations with no thermal photons, namely $\langle\hat{\xi}(t)\rangle = 0$ and $\langle \hat{\xi}(t) \hat{\xi}^{\dagger}(t')\rangle = \delta(t-t')$ with other two-point correlators equal to zero. Under these assumptions, the observable's variance can be written as (see Appendix~\ref{app_InputOutputHomodyne} for details),
\begin{equation}
\label{eq_NoiseTotal}
(\Delta M(T))^2 = (\Delta M_{\mathrm{spin}}(T))^2+ (\Delta M_{\mathrm{shot}}(T))^2,
\end{equation}
which consists of the intrinsic spin projection noise integrated over time,
\begin{align}
    (\Delta &M_{\mathrm{spin}}(T))^2 = \frac{64\overline{n}}{\kappa}\sum_{j,j'=1}^N \chi_j \chi_{j'}\times\\
    &\int_0^T dt \int_0^T dt' \big[ \langle \hat{s}_j^{z}(t) \hat{s}_{j'}^{z}(t')\rangle - \langle \hat{s}_j^{z}(t)\rangle \langle \hat{s}_{j'}^{z}(t')\rangle\big]\nonumber,
\end{align}
and the integrated shot noise of the drive, which is simply linear in time:
\begin{equation}
\label{eq_ShotNoise}
(\Delta M_{\mathrm{shot}}(T))^2 = T.
\end{equation}
For an initial state that is a separable product state (no initial correlations between spins), the spin noise can be computed to be (see Appendix~\ref{app_SpinDynamics}):
\begin{widetext}
\begin{equation}
\label{eq_Noise}
\begin{aligned}
(\Delta M_{\mathrm{spin}}(T))^2&=\frac{64\overline{n}}{\kappa}\sum_{j=1}^N \frac{\chi_j^2}{\gamma_j^2} \left(\langle \hat{s}_j^{z}(0)\rangle + \frac{1}{2}\right)\left[2\left(1-e^{-\gamma_j T}\right)- 2 e^{-\gamma_j T} \gamma_j T-\left(\langle \hat{s}_j^{z}(0)\rangle + \frac{1}{2}\right)\left(1-e^{-\gamma_j T}\right)^2\right].
\end{aligned}
\end{equation}
\end{widetext}
We have zero noise at zero collection time, $(\Delta M(T))^2 = 0$, since we can be certain that no signal has been detected yet.

%%%%%
\subsection{Resolving the spin-projection noise above photon shot noise}
%%%%%

The noise contributions to the variance are simplest to analyze in the homogeneous parameter regime of constant couplings $\chi_j = \chi$ and decay rates $\gamma_j = \gamma$. We will also assume an initial collective product state of all spins residing along the equator of the Bloch sphere, $\langle\hat{s}_j^z(0)\rangle= 0$ (see Fig.~\ref{fig:fig1}(c)). Such a state is typical for standard Ramsey spectroscopy protocols; later, we will consider a signal which is an infinitesimal perturbation away from this initial state. Such a state is also ideal for studying spin-projection-noise properties of the system, since it has maximal variance in the spin polarization observable. The spin-projection noise for homogeneous parameters is
\begin{equation}
\begin{aligned}
(\Delta M_{\mathrm{spin}}(T))^2 = \frac{16\overline{n} N \chi^2}{\kappa \gamma^2}\big[\left(1-e^{-\gamma T}\right)&\left(3+e^{-\gamma T}\right)\\
&-4 e^{-\gamma T}\gamma T\big].
\end{aligned}
\end{equation}
Fig.~\ref{fig:SignalAndNoise}(a) shows this noise and the shot noise as a function of collection time $T$. At short collection time $\gamma T \ll 1$ the spin-projection noise is quadratic in $T$:
\begin{equation}
\begin{aligned}
(\Delta M_{\mathrm{spin}}(T))^2=\frac{16 \overline{n} N\chi^2}{\kappa} T^2 + \mathcal{O}\left(\gamma^3 T^3\right).
\end{aligned}
\end{equation}
This is simply proportional to the initial variance of the spin ensemble times a quadratic time growth $\chi^2 T^2$. The spin noise grows monotonically, then saturates on a timescale of roughly $T \sim 1/\gamma$, as the signal is decaying at rate $\gamma$ and the correlator only picks up significant contributions from short times. The final saturation value at infinitely long collection time is:
\begin{equation}
\lim_{T \to \infty} (\Delta M_{\mathrm{spin}}(T))^2 = \kappa\frac{48\overline{n} N\chi^2}{\gamma^2}.
\end{equation}
In this limit, the spins have finished decaying down to their ground state and do not contribute any further noise.

We can see from Fig.~\ref{fig:SignalAndNoise}(a) that for sufficiently strong driving (i.e. large resonator photon number $\overline{n}$), there is a range of collection times $T$ for which the spin noise dominates the shot noise,
\begin{equation}
(\Delta M_{\mathrm{spin}}(T))^2 \geq (\Delta M_{\mathrm{shot}}(T))^2.
\end{equation}
We are interested in determining when this condition can be achieved parametrically. Still assuming homogeneous parameters, and inserting the short-time expansion for $(\Delta M_{\mathrm{spin}}(T))^2$, we obtain:
\begin{equation}
\frac{(\Delta M_{\mathrm{spin}}(T))^2}{(\Delta M_{\mathrm{shot}}(T))^2} = \frac{16\overline{n} N\chi^2 T}{\kappa} \geq 1.
\end{equation}
As a rough estimate, we can set the collection time needed to pick up all available signal to $T = 1/\gamma$; going further will provide little benefit as the spins will have decayed. Inserting this value gives us a parametric condition:
\begin{equation}
\label{eq_FinalSmallParameter}
\lambda  = \frac{16 \chi^2\overline{n} N}{\kappa \gamma}=\frac{16g^4\overline{n}N}{\Delta^2 \kappa \gamma} \geq 1,
\end{equation}
where $\lambda$ characterizes the quality of the system's measurement resolution. If this condition is satisfied, we can expect the dispersive homodyne measurement to be spin-projection noise resolved. Satisfying $\lambda \geq 1$ is also a necessary precursor to using entangled spin-squeezed states for more precise measurements with this protocol, as will be shown in Section~\ref{sec:Entanglement}. Note that, while this parameter implies one can always do better with a smaller detuning $\Delta$, one must still satisfy other approximations such as remaining in the dispersive regime; a thorough analysis of parametric dependence is provided in the next subsection.

Fig.~\ref{fig:SignalAndNoise}(b) plots the total measurement variance, normalized by the shot noise $(\Delta M_{\mathrm{shot}}(T))^2 = T$. We see that for sufficiently large $\lambda \gg 1$, there is a characteristic noise `bump' above the shot-noise background that occurs due to the presence of the spins. While we will study a specific signal to measure in the next section, a first experimental implementation may seek to first prove that it is spin-projection-noise limited by measuring the presence of such a `bump'. While the same increase in noise could also be observed for, e.g., an infinite-temperature thermal state as opposed to a coherent superposition, one can easily distinguish the two by rotating the initial state by $\pi/2$ from the $x$ to the $z$ axis of the Bloch sphere. A pure initial state would then see no extra noise at short times (if it now starts in an eigenstate of $\hat{s}_j^z$ with zero variance), while a thermal state would retain the same noise profile.

%%%%%
\subsection{Parametric dependence and approximations}
\label{sec:Approximations}
%%%%%

Before moving on, we review the approximations made thus far. The dispersive approximation required $|g_j| \sqrt{\overline{n}}\ll |\Delta-\delta_j|$ for all spins $j$. For our considerations of a collective signal, we want this to be true for the \textit{average} spin -- there can be some exceptions such as near-resonant spins but, provided they are few in number, their contribution should be negligible. For homogeneous $g_j = g$ and assuming $|\Delta| \gtrsim |\delta_j|$ (allowing us to roughly approximate $|\Delta-\delta_j| \approx |\Delta|$ for an average spin), we must have $g \sqrt{\overline{n}} \ll |\Delta|$.

The homodyne current depending on the spin polarization $\sim \sum_j \chi_j\langle\hat{s}_j^{z}\rangle$ assumes the resonator dynamics on timescale $\sim \kappa$ are much faster than any dynamics involving the spins; hence for homogeneous parameters $\chi_j = \chi, \gamma_j = \gamma$ we must have $\gamma, \chi \ll \kappa$.

In deriving the dispersive model, we drop cross terms in both the Hamiltonian and the dissipators (see Appendix~\ref{app_SchriefferWolff}), which requires $|g_{j}g_{j'}|/|\Delta-\delta_j| \ll |\delta_j - \delta_{j'}|$ and $\kappa |g_{j}g_{j'}|/(|\Delta-\delta_j||\Delta-\delta_{j'}|)\ll |\delta_j - \delta_{j'}|$ for all $j \neq j'$. For inhomogeneous broadening width $\sigma_{\delta}$, we can approximate $|\delta_j - \delta_{j'}|\approx \sigma_{\delta}$ for the average spin. If we neglect disorder in $g_j$, the two above conditions become $g^2 \ll\Delta \sigma_{\delta}$, and $\kappa g^2 \ll \Delta^2 \sigma_{\delta}$. Violation of the first condition will enable resonator-mediated spin flip-flop interactions, while violation of the second will cause superradiant enhancement of Purcell decay. Both the resonator detuning $\Delta$ and the broadening width $\sigma_{\delta}$ must be appreciably large for the conditions to hold. The resonator linewidth $\kappa$ must also not be too large for the second condition to hold, but we do not strictly require a broader spin distribution than resonator decay rate $\kappa < \sigma_{\delta}$, as the conditions can still be satisfied by increasing the resonator detuning $\Delta$. For parameters typical to state-of-the-art experimental implementations (see Table~\ref{table_Params}), all of the above approximations are very well satisfied for spin-resonator detuning $\Delta \gtrsim \sigma_{\delta}$. Generally, the challenge is achieving a high enough spin-resonator coupling $g$.

Having reviewed the requisite approximations in deriving our measurement quality parameter, we now further analyze its explicit form. First, we can write $\lambda$ in terms of the collective cooperativity $\mathcal{C}=4Ng^2/(\kappa \gamma)$,
\begin{equation}
\lambda = \frac{4\mathcal{C}g^2 \overline{n}}{\Delta^2}.
\end{equation}
This form implies one can always be more sensitive to spin-projection noise by reducing the spin-resonator detuning $\Delta$. In principle, this behavior will break down once we approach $\Delta = g\sqrt{\overline{n}}$ and the dispersive approximation fails as described above. Thus, replacing $\Delta$ by $g \sqrt{\overline{n}}$ yields a (roughly) maximized measurement quality parameter  over resonator detunings $\lambda_{\mathrm{max}} = 4\mathcal{C}$, i.e., just the collective cooperativity up to a prefactor.

However, as we will explore in Section~\ref{subsec_Inhomogeneity}, the effectiveness of measurement will degrade long before we approach $\Delta = g \sqrt{\overline{n}}$. Due to inhomogeneous broadening, different spins will start to undergo resonator-mediated Purcell decay at different rates as soon as the resonator detuning is comparable to the spin frequency distribution width, $\Delta \lesssim \sigma_{\delta}$. We still want to optimize $\Delta$ to be as small as possible, but its minimum feasible value is now set by $\sigma_{\delta}$, below which it will become difficult to measure a signal. If we thus set $\Delta = \sigma_{\delta}$, we find a different measurement quality parameter maximized over $\Delta$:
\begin{equation}
\begin{aligned}
\lambda_{\mathrm{max}} &=  4\mathcal{C}\mathcal{C}_{\mathrm{inh}},\\
\mathcal{C}_{\mathrm{inh}} &= \frac{\overline{n} g^2}{\sigma_{\delta}^2},
\end{aligned}
\end{equation}
where $\mathcal{C}_{\mathrm{inh}}$ characterizes the quality of the spin-resonator couplings relative to the inhomogeneous broadening. Note that we still require the dispersive approximation to be valid, hence $g^2 \overline{n} \ll \Delta^2 \approx \sigma_{\delta}^2$, meaning the numerator of $\mathcal{C}_{\mathrm{inh}}$ must be much smaller than the denominator and hence $\mathcal{C}_{\mathrm{inh}} \leq 1$.

Combining the value of the optimized measurement-quality parameter limited by inhomogeneous broadening (valid for ensembles with a frequency distribution much broader than the spin-resonator couplings $g \sqrt{\overline{n}}$) and the prior value limited just by the dispersive approximation, we find an overall parameter maximized over the resonator detuning $\Delta$:
\begin{equation}
\lambda_{\mathrm{max}} = 4\mathcal{C}\times\> \text{min}\left(1,\mathcal{C}_{\mathrm{inh}}\right).
\end{equation}
For a practical measurement seeking to achieve, e.g., a good signal-to-noise ratio for a Ramsey-style experiment, as will be studied in Section~\ref{subsec_Inhomogeneity}, one must satisfy $\lambda_{\mathrm{max}} \geq 1$. For this to be true, the collective cooperativity $\mathcal{C}$ must compensate for the effect of broadening described by $\mathcal{C}_{\mathrm{inh}}$ (since the latter is always less than 1).

%%%%%
\subsection{Benchmarking other sources of experimental noise}
%%%%%

We now analyze the impact of other possible imperfections on our proposed measurement schemes.

One major source of error is \textit{phase noise}, present either in the resonator itself or in the laser drive. We provide a detailed analysis of such noise on our protocol in Appendix~\ref{app_PhaseNoise}. The main result is that our condition for resolving intrinsic spin noise gains an additional factor,
\begin{equation}
\lambda \to \frac{\lambda}{1 + \frac{32 \Gamma_L \overline{n}}{\kappa}},
\end{equation}
where $\Gamma_L$ is the width of the phase-noise spectrum, assuming it is of Lorentzian shape~\cite{rabl2009phase}. We see that, in the presence of such phase noise, the scaling improvement with resonator photon number $\lambda \sim \overline{n}$ is eventually lost, although for state-of-the-art platforms $\Gamma_L$ can nonetheless be small enough to still allow $\lambda \geq 1$ to be satisfied~\cite{gao2007noise}.

Another possible source of error is a finite number of experimental runs. If we only run the experiment $N_{\mathrm{runs}}$ times, the resulting variance in outcomes will be a sample variance that may be different from the true $(\Delta M(T))^2$. We study the effects of this finite sampling in Appendix~\ref{app_FiniteRuns}. For homogeneous parameters, resolving spin projection noise requires the following minimum number of runs:
\begin{equation}
N_{\mathrm{runs}} \geq \left(1 + \frac{1}{\lambda}\frac{ \gamma T}{\left(1-e^{-\gamma T}\right)\left(3+e^{-\gamma T}\right)-4e^{-\gamma T}\gamma T}\right)^2.
\end{equation}
In the limit of a perfect system $\lambda \to \infty$, only one experimental run is needed. Various series expansions of this condition are detailed in Appendix~\ref{app_FiniteRuns}.

One more common experimental pitfall is finite measurement efficiency $\eta \leq 1$, for which the observed signal reads,
\begin{equation}
\hat{I}_{\mathrm{out}}(t) = \sqrt{\eta} \left[\hat{a}_{\mathrm{out}}(t) +\hat{a}_{\mathrm{out}}^{\dagger}(t)\right] + \sqrt{1-\eta}\>\hat{x}_{\mathrm{noise}}(t),
\end{equation}
where $\hat{x}_{\mathrm{noise}}(t) = \hat{\xi}_{\mathrm{noise}}(t) + \hat{\xi}_{\mathrm{noise}}^{\dagger}(t)$, and $\hat{\xi}_{\mathrm{noise}}(t)$ is white-noise distributed (expectation values $\langle\hat{\xi}_{\mathrm{noise}}(t)\rangle = 0$, $\langle\hat{\xi}_{\mathrm{noise}}(t)\hat{\xi}_{\mathrm{noise}}^{\dagger}(t')\rangle = \delta(t-t')$ with all other correlators zero). Thus, a fraction $1-\eta$ of the signal is essentially replaced by white noise. It is straightforward to show that non-unit homodyne efficiency reduces the time-integrated spin noise by $(\Delta M_{\mathrm{spin}}(T))^2 \to \eta (\Delta M_{\mathrm{spin}}(T))^2$, while the shot-noise remains unchanged at $(\Delta M_{\mathrm{shot}}(T))^2 = T$. The measurement quality parameter thus inherits the efficiency factor,
\begin{equation}
\lambda \to \lambda \eta.
\end{equation}

%%%%%%
\section{Resonator-based spin-ensemble magnetometry}
%%%%%

Having analyzed the noise properties of our setup, we now turn to the ultimate application of interest: the ability to sense a small rotation of the spin ensemble with SQL or even better-than-SQL precision.  
Suppose all spins still start in a collective product state, but infinitesimally rotated away from the equator of the Bloch sphere, $\langle\hat{s}_j^{z}(0)\rangle = \sin(\theta)/2$ with $\theta \ll 1$ (see Fig.~\ref{fig:fig1}(c)). We seek to extract information about $\theta$  by measuring the change in spin polarization via the resonator. As it is standard in Ramsey spectroscopy, the phase $\theta$ is assumed to be infinitesimally small, hence we consider the limit $\theta \to 0$ in all our analysis hereafter. For solid state ensembles, the spin polarization is typically measured via fluorescence or absorption measurements of the sensors; here we will use the homodyne current coming from the resonator.

Before going into quantitative detail, we overview the anticipated results. For a perfect measurement limited by spin projection noise only, without any additional noise sources or decoherence processes, we can anticipate that Ramsey spectroscopy measurements of infinitesimally small $\theta$ will have a signal-to-noise ratio (SNR) of $\sqrt{N}$ for non-entangled spins (which is the case we consider). Our goal is to understand how close we can come to this limit using the resonator-based dispersive readout.

The SNR of the measurement will non-trivially depend on the homodyne-current collection time $T$. 
Since $(\Delta M_{\mathrm{spin}}(T))^2$ grows faster than the shot-noise term $(\Delta M_{\mathrm{shot}}(T))^2$, integrating for long times $T$ suppresses the impact of shot noise and ensures spin-projection noise is dominant. 
However, integrating much longer than the $T_1$ time, i.e., $\gamma T \gg 1$, will not improve the SNR because the spins will have decayed and there will be no signal any more.
There will thus be an optimal $T$ that we will describe quantitatively below.  

%%%%%%
\subsection{Signal-to-noise ratio}
%%%%%

The signal $\langle\hat{M}(T)\rangle$ is proportional to the time integral $\sim \int_0^{T} dt\langle\hat{s}_j^z(t)\rangle$ and hence picks up contributions from any $\langle\hat{s}_{j}^z(t)\rangle \neq 0$. Since the spins relax towards $\langle\hat{s}_{j}^z(t \to \infty)\rangle = -1/2$, there will still be a signal even after the spins have finished decaying, albeit not one that provides information on $\theta$. If we want the measurement to encode only the initial condition of the spins, it can be useful to consider a differential signal relative to some 'reference' time-evolution $\langle\hat{s}_{j,0}^{z}(t)\rangle$,
\begin{equation}
\langle\hat{M}_{\mathrm{signal}}(T)\rangle = \frac{8\sqrt{\overline{n}}}{\sqrt{\kappa}} \sum_j \chi_j \int_{0}^{T} dt \left[\langle\hat{s}_{j}^{z}(t)\rangle-\langle\hat{s}_{j,0}^{z}(t)\rangle\right].
\end{equation}
For example, we can consider $\langle\hat{s}_{j,0}^{z}(t)\rangle$ to be the time-evolution for spins starting on the equator of the Bloch sphere $\langle\hat{s}_{j,0}^{z}(0)\rangle = 0$ as in the prior section, and seek to measure the difference in signal that is due to the infinitesimal rotation $\langle\hat{s}_{j}^{z}(0)\rangle = \sin(\theta)/2$.

The signal-to-noise ratio for measuring $\theta$ can be written as,
\begin{equation}
\label{eq_SNRfull}
\mathrm{SNR}(T) = \frac{\left|\frac{\partial}{\partial \theta}\langle\hat{M}_{\mathrm{signal}}(T)\rangle\right|}{|\Delta M(T)|},
\end{equation}
We assume that the noise in the denominator is coming only from the signal with $\theta \neq 0$, meaning that the reference signal (e.g. with $\theta = 0$) is already known and calibrated. Since we assume an infinitesimally small $\theta$, both the numerator and denominator above are evaluated for $\theta \to 0$.

For homogeneous parameters we can write an explicit expression for this SNR:
\small
\begin{equation}
\label{eq_SNR}
\frac{\mathrm{SNR}(T)}{\sqrt{N}} = \frac{1-e^{-\gamma T}}{\sqrt{\left(1-e^{-\gamma T}\right)\left(3+e^{-\gamma T}\right)-4e^{-\gamma T}\gamma T+\frac{1}{\lambda} \gamma T}},
\end{equation}
\normalsize
where $\lambda=16 \chi^2 \overline{n} N /(\kappa \gamma)$ is our measurement quality factor from before.
In Fig.~\ref{fig:SNR}(a) we plot this SNR relative to the standard quantum limit $\mathrm{SNR} = \sqrt{N}$. One can approach $\mathrm{SNR}(T)/\sqrt{N} = 1$ for a perfect system with $\lambda \to \infty$. The optimal integration time occurs around $\gamma T \approx 1$ for poorer-quality systems, but shifts to shorter times for higher-quality ones since the signal can be acquired very quickly.

We can series expand the expression; for short times $\gamma T \ll 1$ and finite $\lambda$, it reads:
\begin{equation}
\frac{\mathrm{SNR}(T)}{\sqrt{N}}= \sqrt{\lambda \gamma T}+\mathcal{O}\left(\lambda^{3/2}\gamma^{3/2} T^{3/2}\right).
\end{equation}
If we are in the limit of an infinitely high-quality system $\lambda \to \infty$, the expression expands at short times to:
\begin{equation}
\text{lim}_{\lambda \to \infty}\frac{\mathrm{SNR}(T)}{\sqrt{N}}= 1 - \frac{\gamma T}{3} + \mathcal{O}\left(\gamma^2 T^2\right).
\end{equation}
For $\lambda \gtrsim 1$, we can also find the optimum of the full expression to be at approximately $\gamma T = \sqrt{3/(2\lambda)}$. At this maximum, we have $\mathrm{SNR}\approx 1-\sqrt{2/(3\lambda)}$.

%%%%%
\subsection{Inhomogeneity effects}
\label{subsec_Inhomogeneity}
%%%%%

\begin{figure*}
    \center
    \includegraphics[width=\linewidth]{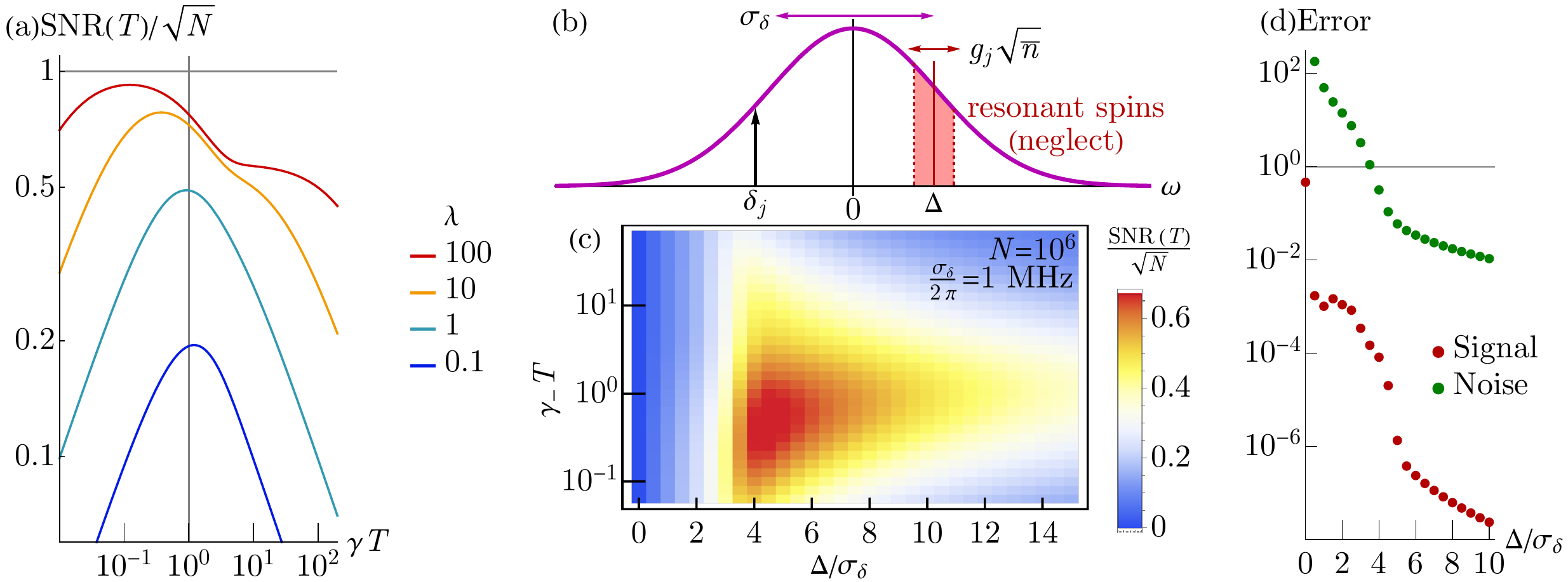}
    \caption{(a) SNR of the measurement relative to the standard quantum limit $\sqrt{N}$ from Eq.~\eqref{eq_SNR}, assuming an infintesimally small angle $\theta \ll 1$ and homogeneous system parameters. (b) Frequency schematic for an inhomogeneous ensemble. We still use the parameters from Table~\ref{table_Params}, still fix spin-resonator couplings at $g_j/(2\pi) = 50$ Hz, but randomly sample non-uniform spin frequencies $\delta_j$ from a Gaussian distribution with standard deviation $\sigma_{\delta} = 2\pi \times 1$ MHz, leading to non-uniform decay rates $\gamma_j = \gamma_{-} + \kappa |g_j|^2/(\Delta-\delta_j)^2$ and dispersive couplings $\chi_j=|g_j|^2/(\Delta-\delta_j)$. Since this sampling can lead to some spins breaking the dispersive approximation, we discard all spins with sampled frequencies satisfying $|\Delta - \delta_j|\leq g \sqrt{\overline{n}}$ (highlighted red region). (c) SNR for an inhomogeneous ensemble using the above approach. The SNR is calculated for 10 realizations of random spin frequencies $\delta_j$ and averaged together. (d) Difference in signal and spin noise between inhomogeneous and homogeneous parameters from Eq.~\eqref{eq_InhomogeneousNoiseComparison} for a collection time $\gamma_{-} T = 1$.}
    \label{fig:SNR}
\end{figure*}

So far, we assumed homogeneous couplings $\chi_j = \chi$ and decay rates $\gamma_j = \gamma$. 
Under this assumption, one can always improve SNR by decreasing the resonator detuning $\Delta$ as long as the dispersive approximation holds. 
However, as discussed in Sec.~\ref{sec:Approximations}, a smaller resonator detuning $\Delta$ ceases to be useful even in the dispersive regime once inhomogeneities are factored in.
In practice, for hybrid-circuit-QED platforms, $\chi_j$ can be very inhomogeneous due to either broadening of spin detunings $\delta_j$ or unequal couplings $g_j$ to the resonator. 

We now explicitly evaluate the effects of inhomogeneity numerically, using the candidate experimental system broadly described in Section~\ref{sec_expt}. We use the typical parameters from Table~\ref{table_Params}, with spin frequencies $\delta_j$ randomly sampled from a Gaussian distribution with standard deviation $\sigma_{\delta}$. 
Fig.~\ref{fig:SNR}(b) shows a schematic of the spin frequency range. 
We then calculate both the signal [Eq.~\eqref{eq_Signal}] and the noise [Eq.~\eqref{eq_ShotNoise} plus Eq.~\eqref{eq_Noise}] for dispersive couplings $\chi_j = |g_j|^2/(\Delta - \delta_j)$ and decay rates $\gamma_j = \gamma_{-} + \kappa |g_j|^2/(\Delta - \delta_j)^2$ by performing the underlying sums over $j$. Since a sufficiently small $\Delta/\sigma_{\delta}\lesssim 1$ can lead to some spins being near-resonant and the dispersive approximation breaking down, we discard spins $j$ that have $|\Delta-\delta_j| \leq g_j\sqrt{\overline{n}}$ from the calculations;
this is consistent with the expectation that such spins would decay very quickly, and not contribute to the measurement signal.  For all parameters used, more than $99 \%$ of this spins in the ensemble are retained under this approximation. Fig.~\ref{fig:SNR}(c) shows the resulting SNR as a function of both collection time $T$ and resonator detuning $\Delta$. The calculation is repeated for 10 random realizations of spin frequencies, and the plotted $\mathrm{SNR}$ is an average over these realizations. The spin-resonator couplings are still taken to be constant, $g_j = g$ for simplicity, although in practice they can also be inhomogeneous and create further disorder.

Figure~\ref{fig:SNR}(c) confirms that a too large resonator detuning $\Delta \to \infty$ is unfavorable for the SNR because the dispersive couplings vanish, $\chi_j \to 0$, while the decay rates remain finite, $\gamma_j \geq \gamma_-$. 
As a consequence, no information can be gained before the spins decay.

In the opposite regime of a small detuning $\Delta$, we find somewhat surprisingly that the SNR drops dramatically and rather sharply as $\Delta/\sigma_\delta$ is decreased below an order-1 threshold value. 
This is due to the fact that disorder leads to different decay rates for different spins.
Consider first the simpler case of a homogeneous ensemble. As shown in Fig.~\ref{fig:SNR}(a), 
in this case there is an optimal integration time $T$, with the sharp maximum here related to the fact that all spins decay on the same timescale. 
Returning to a disordered ensemble, while inhomogeneous couplings $\chi_j$ simply lead to a renormalization of the overall prefactor $\sum_j \chi_j^2$ in the spin noise (see Eq.~\eqref{eq_Noise}), sufficiently large spin-to-spin variation of the decay rates 
$\gamma_j$ leads to something more severe:  different spins lose information in their initial z-polarization on very different timescales. 
If the spins' detunings vary by a mean value of $\vert \delta_j - \delta_{j'}\vert \sim \sigma_{\delta}$, the decay rates $\gamma_j$ will cease to be homogeneous for resonator detuning $\Delta \sim \sigma_{\delta}$, with a proportionality constant of order unity. 
For fixed $T$ (e.g., the optimal integration time in a homogeneous system), many spins will have missed their window of optimal signal acquisition, while others have not reached it yet, leading to sub-optimal signal, while spin-projection noise continues to accrue for all spins regardless, causing the breakdown of the SNR. 
We emphasize that such a breakdown occurs much earlier than potential failure of the dispersive approximation -- indeed, even for a resonator with no detuning, $\Delta = 0$, for sufficiently large spin frequency broadening $\sigma_{\delta}$, most spins will still remain dispersive.

To highlight the impact of inhomogeneous system parameters on the SNR more clearly, we analyze the roles of signal and noise separately.
As discussed previously, inhomogeneity implies that the observable $\sim \sum_j\chi_j \langle\hat{s}_j^z (t)\rangle$ depends not on the true collective polarization $\langle\hat{S}^{z}\rangle = \sum_j \langle\hat{s}_j^{z}\rangle$, but rather a weighted sum. 
A common approach in the field is to simply account for this effect by treating inhomogeneous ensembles as homogeneous ones with a reduced effective atom number~\cite{hu2015entangled}, which, however, neglects the different decay timescales.
In our case, we can compute the difference in both the signal and the noise compared to averaged homogeneous couplings:
\begin{equation}
\label{eq_InhomogeneousNoiseComparison}
\begin{aligned}
\text{Error}_{\mathrm{signal}} = \left|\frac{\langle \hat{M}(T)\rangle}{\langle \hat{M}_{\mathrm{homog}}(T)\rangle}-1\right|,\\
\text{Error}_{\mathrm{spin\> noise}} = \left|\frac{(\Delta M_{\mathrm{spin}}(T))^2}{(\Delta M_{\mathrm{spin,homog}}(T))^2}-1\right|.
\end{aligned}
\end{equation}
Here $\langle\hat{M}_{\mathrm{homog}}(T)\rangle$ and $(\Delta M_{\mathrm{spin,homog}}(T))^2$ are $\langle\hat{M}(T)\rangle$ and $(\Delta M_{\mathrm{spin}}(T))^2$, but setting all $\chi_j \equiv \overline{\chi}$, $\gamma_j \equiv \overline{\gamma}$, with averaged $\overline{\chi} = \frac{1}{N}\sum_j \chi_j$, $\overline{\gamma} = \frac{1}{N} \sum_{j}\gamma_j$. In the limit of perfectly homogeneous parameters, both errors go to zero. Fig.~\ref{fig:SNR}(d) shows these errors for the same Gaussian-distributed spin detunings. As the resonator detuning decreases below $\Delta/\sigma_{\delta}\lesssim 1$, the noise for inhomogeneous ensembles becomes orders of magnitude stronger, leading to the decrease in SNR observed for small $\Delta/\sigma_{\delta}$ in Fig.~\ref{fig:SNR}(c). 
While the error of the signal also increases in this regime, the difference is much smaller. 
Note that the reason for the very sharp onset of bad SNR in Fig.~\ref{fig:SNR}(c) at small $\Delta$ is caused by our choice of Gaussian probability distribution for the spin frequencies, which has an exponentially vanishing tail at high frequencies. We abruptly see a strong effect once the resonator detuning $\Delta$ enters the frequency range where most of the probability distribution resides. Other distributions like Lorentzians will feature a more uniform profile of SNR, though they will still exhibit a characteristic optimum resonator detuning.

%%%%%
\section{Entanglement}
\label{sec:Entanglement}
%%%%%

Our analysis thus far has allowed identification of a regime where spin-projection noise $(\Delta M_{\mathrm{spin}}(T))^2$ is higher than shot noise $(\Delta M_{\mathrm{shot}}(T))^2$. The former can be reduced with an entangled initial state of the spins, such as a spin-squeezed state with smaller variance of the spin polarization. While the variance is reduced at the start of measurement, the reduction will disappear throughout the collection time as the spins decay. It is thus worth asking how the noise behaves for an initially entangled sensing state. Here, we perform a general analysis, assuming homogeneous system parameters, to see whether entanglement can be resolved.

\begin{figure}
    \center
\includegraphics[width=0.7\linewidth]{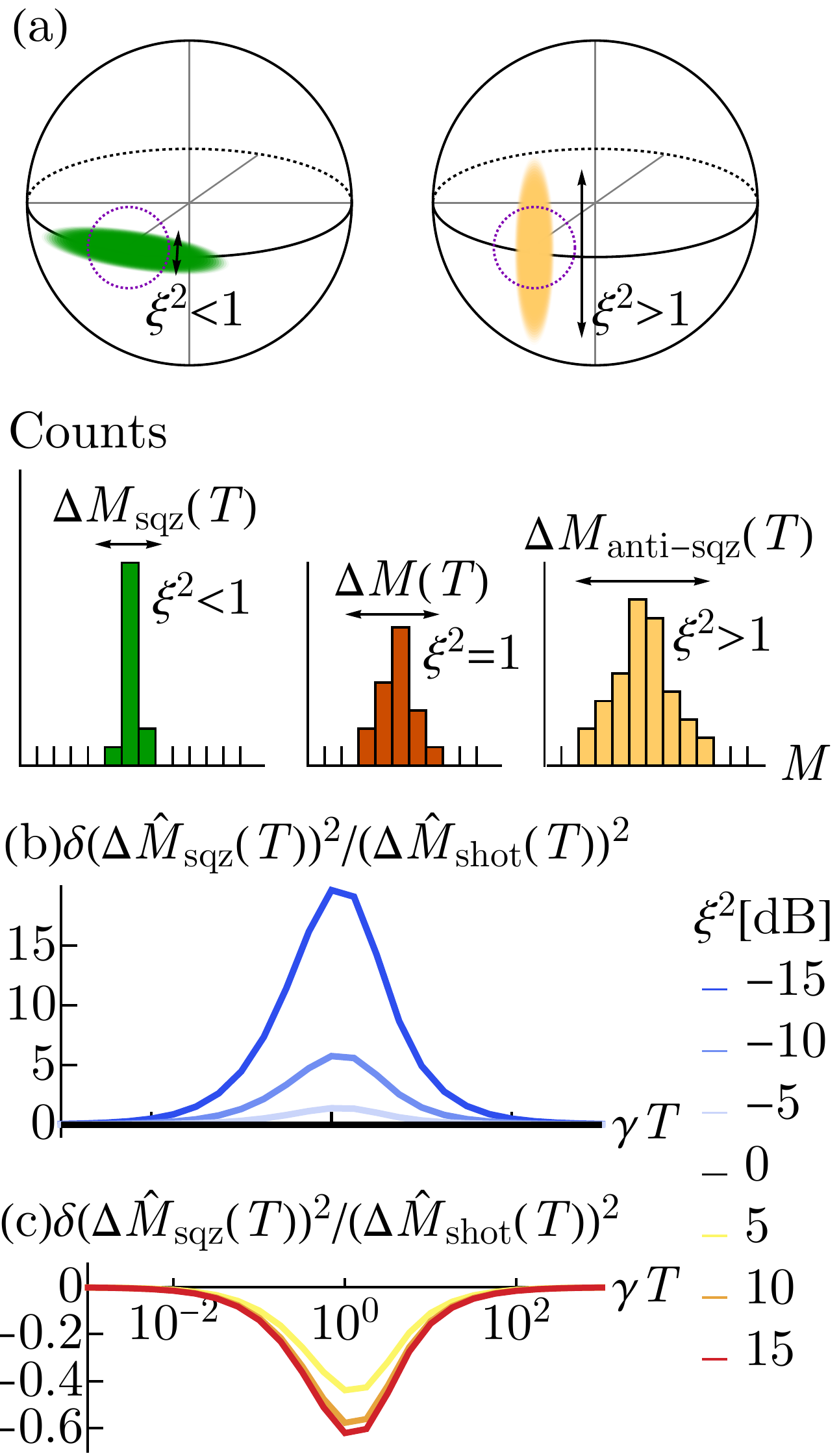}
    \caption{(a) Schematic of an entangled initial spin-squeezed state characterized by Wineland parameter $\xi^2$. The subsequent variance in the integrated homodyne current will shrink (grow) for (anti-)squeezed states with $\xi^2 <1$ ($\xi^2 > 1$) for fixed collection time $T$. (b-c) Noise difference (change in variance) of the homodyne measurement from Eq.~\eqref{eq_spinNoiseEntangled}, normalized by the shot noise $(\Delta M_{\mathrm{shot}}(T))^2 = T$. We consider both (b) anti-squeezed and (c) squeezed states. The squeezing is reported in decibels $\xi^2[\text{dB}] = -10 \text{log}_{10}(\xi^2)$. We use homogeneous system parameters from Table~\ref{table_Params} except for the spin-resonator coupling, which we fix to $g_j = 2\pi \times 63$ Hz, thereby setting the measurement quality parameter to $\lambda = 10$.}
    \label{fig:Entanglement}
\end{figure}

We still consider a Ramsey-style sensing state pointing along the positive $x$-axis of the Bloch sphere, i.e., $\langle\hat{S}^{x}(0)\rangle = N/2$.
Using this state, we sense perturbations along $\pm z$ by some infinitesimal angle $\theta \ll 1$. As depicted in Fig.~\ref{fig:Entanglement}(a), this sensing state's phase-space distribution is initially squeezed before the measurement is performed, e.g., via an entangling Hamiltonian such as the all-to-all one-axis twisting (OAT) interaction $\hat{H}_{\mathrm{OAT}} = \hat{S}^{z}\hat{S}^{z}$. Real solid-state implementations will almost certainly have a more complex and shorter-range Hamiltonian; we study OAT for simplicity. 
Following the generation of the entangled sensing state, the state is rotated about the $x$ axis such that the axis with reduced variance is aligned with the $\pm z$ direction, with $\langle(\hat{S}^{z})^2(0)\rangle \leq N/4$. The improved sensitivity of this state is determined by the Wineland squeezing parameter $\xi^2 = N \langle (\hat{S}^{z})^2(0)\rangle/\langle\hat{S}^{x}(0)\rangle^2$. We consider weakly entangled states far from the Heisenberg limit of $\xi^2 = 1/N$, for which the spin contrast $\langle\hat{S}^{x}(0)\rangle=N/2$ does not appreciably decay under the OAT Hamiltonian, thus simplifying the Wineland parameter to just be proportional to the initial variance $\xi^2 = 4 \langle (\hat{S}^{z})^2(0)\rangle/N$. The state is squeezed when $\xi^2 <1$, and anti-squeezed for $\xi^2 > 1$; the magnitude of (anti-)squeezing is characterized by $\xi^2-1$.

The initial entanglement changes the variance of the homodyne measurement to $(\Delta M_{\mathrm{sqz}}(T))^2$ (for $\xi^2<1$) or $(\Delta M_{\mathrm{anti-sqz}}(T))^2$ (for $\xi^2 > 1$), which is given by (see Appendix~\ref{app_Entanglement} for details):
\begin{equation}
\label{eq_spinNoiseEntangled}
\begin{aligned}
(\Delta M_{\mathrm{(anti-)sqz}}(T))^2 &= (\Delta M(T))^2  \\
&+\delta (\Delta M_{\mathrm{(anti-)sqz}}(T))^2,\\
\delta (\Delta M_{\mathrm{(anti-)sqz}}(T))^2 &=  \frac{\lambda}{ \gamma}\left(\xi^2-1\right)\left(1-e^{-\gamma T}\right)^2,\\
\end{aligned}
\end{equation}
The change in noise $\delta (\Delta M_{\mathrm{(anti-)sqz}}(T))^2$ is proportional to our measurement-quality parameter from Eq.~\eqref{eq_FinalSmallParameter}, as well as the squeezing strength $\xi^2-1$. A squeezed initial state with $\xi^2 < 1$ reduces the overall noise, whereas an anti-squeezed state $\xi^2 > 1$ would increase it instead. 

To experimentally verify the preparation of an entangled spin-squeezed state, one can perform an experiment measuring the quantum noise of the state with different angles of rotation for its phase-space distribution. Finding a decreased variance for the specific angles at which the squeezed axis aligns with the $\pm z$ direction, and an increased variance if the anti-squeezed axis is aligned, provides a robust demonstration of entanglement. One would thus seek to see a difference in variance relative to a non-entangled state. Figs.~\ref{fig:Entanglement}(b-c) plot the change in variance normalized by the shot noise background due to an anti-squeezed or squeezed state respectively. The latter is much easier to observe, though it does not directly constitute proof of entanglement on its own. The normalized difference exhibits a maximum around $\gamma T \approx 1.25$ in both cases, although such a timescale can be quite long for solid-state spins with slow $T_1$ decay rate, so one may instead use as long a collection time as is experimentally viable.

Note that as for non-entangled initial states, one needs a minimum number of experimental runs $N_{\mathrm{runs}}$ to resolve a change in variance due to entanglement. Parametric conditions for $N_{\mathrm{runs}}$ are provided in Appendix~\ref{app_Entanglement}.

%%%%%
\section{Conclusions and outlook}
%%%%%

We have shown that by using dispersive coupling to a resonator, one can measure the spin polarization of an ensemble of solid-state spins via homodyne detection with spin-projection-limited noise provided the measurement quality parameter from Eq.~\eqref{eq_FinalSmallParameter} satisfies $\lambda \geq 1$. For a fixed collection time $T$, we can also explicitly determine the total measurement variance in the presence of an initially entangled state [Eq.~\eqref{eq_spinNoiseEntangled}, also valid for non-entangled states by setting $\xi^2 = 1$]. Notably, our results apply to both squeezed and anti-squeezed states.

One future outlook is to consider non-detuned resonators, for which the dispersive approximation fails and spins decay directly at coupling rate $g_j$. While for strong inhomogeneous broadening most spins will still couple dispersively, one can engineer experiments that focus on operating with the resonant spins specifically. The ensuing signal from resonant spins could in theory be collected much more quickly, perhaps even with superradiant enhancement, although the theoretical treatment becomes more complicated since the spin polarization observable $\hat{s}_j^z$ no longer decouples from the resonator dynamics.

Another future avenue of research is to incorporate spin flip-flop interactions, either direct dipole-dipole or resonator-mediated. Strong broadening suppresses such interactions, and the analysis in this work neglects them. Nonetheless, for very dense ensembles in frequency space, the effect of flip-flops may become non-negligible, causing diffusion of excitations between spins with different spin-resonator couplings $g_j$ and thus changing the measured signal. One possible approach is to coarse-grain frequency space, reducing the full ensemble to a smaller set of larger spins living in discrete frequency bins, for which semi-classical or cumulant expansion methods can be of use.

\begin{acknowledgments}
We thank Patrice Bertet, Emily Davis, and Dave Schuster for many helpful discussions. This work was primarily funded by the DOE Q-NEXT Center (Grant No. DOE 1F-60579). M.M. acknowledges funding from the University of Toronto Centre for Quantum Information and Quantum Control (CQIQC). M.K. acknowledges funding from the European Union’s HORIZON Europe research and innovation program via project SPINUS (Grant No. 101135699). J.V. acknowledges support from the NSF QLCI program through grant number OMA-2016245. A.B.J. acknowledges support from the Gordon and Betty Moore Foundation’s EPiQS Initiative via Grant GBMF10279. A.C. acknowledges support from the Simons Foundation through a Simons Investigator award
(Grant No.~00017329).
\end{acknowledgments}

%%%%%%%%%%%%%%%%%%%

\bibliography{NVReadoutBib}
\clearpage
\onecolumngrid
\appendix
\section{Derivation of dispersive model}
\label{app_SchriefferWolff}

This appendix derives the dispersive model in the main text. We start with the bare Hamiltonian, which we split into two pieces,
\begin{equation}
\begin{aligned}
\hat{H} &= \hat{H}_0 + \hat{V},\\
\hat{H}_0 &= \Delta \hat{a}^{\dagger}\hat{a} + \sum_{j=1}^N \delta_j \hat{s}_{j}^{z},\\
\hat{V} &=   \sum_{j=1}^N \left(g_j\hat{a}^{\dagger}\hat{s}_{j}^{-} + g_j^{*}\hat{a} \hat{s}_{j}^{+}\right).
\end{aligned}
\end{equation}
Here, $\hat{H}_0$ is the zeroth-order Hamiltonian representing the largest energy scale in the system, while $\hat{V}$ contains couplings that are block off-diagonal in the basis of $\hat{H}_0$. The Schrieffer-Wolff transformation provides an effective Hamiltonian that is block-diagonal to second order in $\hat{V}$, obtained via a unitary transformation,
\begin{equation}
\label{eq:HSW}
\begin{split}
\hat{H}_{\mathrm{SW}} &= e^{\hat{S}}\hat{H} e^{-\hat{S}} = \hat H + [\hat S, \hat H] + \frac{1}{2!} [\hat S , [\hat S, \hat H]] + \cdots \\
&= \hat{H}_0 + \frac{1}{2}[\hat{S},\hat{V}] + \mathcal{O}\left(||\hat{V}||^3\right),
\end{split}
\end{equation}
where the simplification in the second line is introduced by the Schrieffer-Wolff generator $\hat{S}$ which has been chosen to satisfy
\begin{equation}
\hat{V} + [\hat{S},\hat{H}_0] = 0.
\end{equation}
It is straightforward to check that the above equation is fulfilled by,
\begin{equation}
\hat{S} = \sum_{j=1}^N \frac{1}{\Delta - \delta_j}\left(g_j \hat{a}^{\dagger}\hat{s}_{j}^{-} -g_{j}^{*}\hat{a}\hat{s}_{j}^{+}\right).
\end{equation}
Using this generator, the second-order contribution in the effective Hamiltonian $\hat{H}_\mathrm{SW}$ reads,
\begin{equation}
\begin{aligned}
\frac{1}{2}[\hat{S},\hat{V}] &=\frac{1}{2}\sum_{j,j'=1}^N \frac{1}{\Delta-\delta_j}\left[g_j\hat{a}^{\dagger}\hat{s}_{j}^{-} - g_{j}^{*}\hat{a} \hat{s}_{j}^{+}, g_{j'}\hat{a}^{\dagger}\hat{s}_{j'}^{-} + g_{j'}^{*}\hat{a}\hat{s}_{j'}^{+}\right]\\
&= -2\hat{a}^{\dagger}\hat{a} \sum_{j=1}^N \frac{|g_j|^2}{\Delta - \delta_j}\hat{s}_{j}^{z} -\sum_{j=1}^N \frac{|g_j|^2}{\Delta-\delta_j}\hat{s}_{j}^{z}  - \frac{1}{2}\sum_{j,j'=1}^N \frac{1}{\Delta-\delta_j}\left(g_{j}^{*}g_{j'}\hat{s}_{j}^{+}\hat{s}_{j'}^{-} + g_j g_{j'}^{*}\hat{s}_{j}^{-}\hat{s}_{j'}^{+}\right).
\end{aligned}
\end{equation}
Combining this with the zeroth-order $\hat{H}_0$ gives us the effective Hamiltonian:
\begin{equation}
\label{eq_SWunrotated}
\hat{H}_{\mathrm{SW}} = \Delta \hat{a}^{\dagger}\hat{a} + \sum_{j=1}^N \left(\delta_j  -\frac{|g_j|^2}{\Delta-\delta_j}\right)\hat{s}_{j}^{z} -2\hat{a}^{\dagger}\hat{a} \sum_{j=1}^N \frac{|g_j|^2}{\Delta - \delta_j}\hat{s}_{j}^{z} - \frac{1}{2}\sum_{j,j'=1}^N \frac{1}{\Delta-\delta_j}\left(g_{j}^{*}g_{j'}\hat{s}_{j}^{+}\hat{s}_{j'}^{-} + g_{j} g_{j'}^{*}\hat{s}_{j}^{-}\hat{s}_{j'}^{+}\right).
\end{equation}
Now, we make the observation that the last term is a flip-flop interaction. In a rotating-frame defined by $\hat{H}_0$, we have $\hat{s}_{j}^{-} \to \hat{s}_{j}^{-} e^{i\delta_j t}$, causing matrix elements with $j \neq j'$ to pick up time-dependent phases $\sim e^{i (\delta_j - \delta_j')t}$. If the coupling matrix element $\sim |g_j^{*} g_{j'} /(\Delta -\delta_j)|$ is much smaller than the rotation rate $|\delta_{j} - \delta_{j'}|$, it is rotated out and can thus be neglected. For an inhomogeneously broadened ensemble, $|\delta_{j} - \delta_{j'}|$ is on the order of the ensemble frequency width $\sigma_{\delta}$ on average, meaning \textit{most} spins are far apart in frequency provided $\sigma_{\delta} \gg |g_j^{*} g_{j'} /(\Delta -\delta_j)|$. Assuming we are in this regime, to lowest order we thus drop all flip-flop interactions with $ j \neq j'$. The remaining terms with $j = j'$ are a constant shift $\hat{s}_j^{+} \hat{s}_j^{-} + \hat{s}_j^{-}\hat{s}_j^{+} \sim \mathbbm{1}$ and can also be omitted. The Hamiltonian reduces to
\begin{equation}
\hat{H}_{\mathrm{SW}}=\Delta \hat{a}^{\dagger}\hat{a} + \sum_{j=1}^N \left(\delta_j  -\frac{|g_j|^2}{\Delta-\delta_j}\right)\hat{s}_{j}^{z} -2\hat{a}^{\dagger}\hat{a} \sum_{j=1}^N \frac{|g_j|^2}{\Delta - \delta_j}\hat{s}_{j}^{z} + \mathrm{(const.)}
\end{equation}
If there was no disorder in $g_j$ and $\delta_j$, the flip-flop couplings would not rotate out. The last term in Eq.~\eqref{eq_SWunrotated} would contribute a well-known one-axis twisting (OAT) interaction $\left(\sum_{j=1}^{N} \hat{s}_z^j\right)^2$, which is important for squeezing generation. Here, this effect is suppressed by the disorder. All we have remaining is a single-spin energy shift and the dispersive coupling. We next argue that the perturbative correction to the single-spin shift $\delta_j$ can also be neglected, since it is much smaller and the $\delta_j$ are already random, which simplifies further to:
\begin{equation}
\begin{aligned}
\hat{H}_{\mathrm{SW}} &\approx \Delta \hat{a}^{\dagger}\hat{a} + \sum_{j=1}^N \delta_j \hat{s}_{j}^{z} -2\hat{a}^{\dagger}\hat{a} \sum_{j=1}^N \frac{|g_j|^2}{\Delta - \delta_j}\hat{s}_{j}^{z}.
\end{aligned}
\end{equation}

We can also apply the unitary transformation to the dissipators. The resonator lowering operator becomes,
\begin{equation}
\begin{aligned}
\hat{a} \to e^{\hat{S}} \hat{a} e^{-\hat{S}} &\approx \hat{a} + [\hat{S},\hat{a}]\\
&= \hat{a} + \sum_{j=1}^N \frac{1}{\Delta- \delta_j} [g_j\hat{a}^{\dagger}\hat{s}_{j}^{-} + g_{j}^{*}\hat{a}\hat{s}_{j}^{+},\hat{a}]\\
&= \hat{a} -\sum_{j=1}^N \frac{g_j}{\Delta-\delta_j}\hat{s}_{j}^{-}\\
\end{aligned}
\end{equation}
The resonator loss dissipator takes the form of,
\begin{equation}
\kappa \mathcal{D}[\hat{a}]\rho \to \kappa \mathcal{D} \left[\hat{a} - \sum_{j=1}^N \frac{g_j}{\Delta-\delta_j} \hat{s}_{j}^{-}\right]\rho.
\end{equation}
Expanding this out yields,
\begin{equation}
\begin{aligned}
\kappa \mathcal{D} \left[\hat{a} - \sum_{j=1}^N \frac{g_j}{\Delta-\delta_j} \hat{s}_{j}^{-}\right]\rho &=\kappa \left(\hat{a} \rho \hat{a}^{\dagger}-\frac{1}{2}\hat{a}^{\dagger}\hat{a}\rho - \frac{1}{2}\rho\hat{a}^{\dagger}\hat{a}\right)\\
&-\kappa\sum_{j=1}^N \frac{g_j}{\Delta - \delta_j}\left( \hat{s}_{j}^{-} \rho \hat{a}^{\dagger} -\frac{1}{2}\hat{a}^{\dagger} \hat{s}_{j}^{-}\rho - \frac{1}{2}\rho \hat{a}^{\dagger}\hat{s}_{j}^{-}\right)\\
&- \kappa\sum_{j=1}^N \frac{g_j^{*}}{\Delta - \delta_j}\left( \hat{a} \rho \hat{s}_{j}^{+} -\frac{1}{2}\hat{s}_{j}^{+}\hat{a}\rho - \frac{1}{2}\rho\hat{s}_{j}^{+} \hat{a}\right)\\
&+\kappa\sum_{j,j'=1}^N \frac{g_j g_{j'}^{*}}{(\Delta - \delta_j) (\Delta - \delta_{j'})}\left(\hat{s}_{j}^{-}\rho \hat{s}_{j'}^{+} - \frac{1}{2}\hat{s}_{j;}^{+}\hat{s}_{j}^{-}\rho - \frac{1}{2}\rho\hat{s}_{j'}^{+}\hat{s}_{j}^{-}\right).
\end{aligned}
\end{equation}
We once again observe that in a rotating-frame defined by $\hat{H}_0$, we have $\hat{a}\to \hat{a}e^{i\Delta t}$ and $\hat{s}_{j}^{-} \to \hat{s}_{j}^{-} e^{i\delta_j t}$. The cross terms (second and third line) pick up phases $\sim e^{i(\Delta -\delta_j)t}$, and thus rotate out provided $|\kappa g_j/(\Delta-\delta_j)| \ll |\Delta - \delta_j|$, which is generically true for large enough resonator detuning $\Delta$. The last term picks up a phase $\sim e^{i(\delta_j - \delta_{j'})t}$, which as before is comparable to the broadening width $\sigma_{\delta}$ on average. Provided this broadening width is much larger than the matrix element $|\kappa g_{j} g_{j'}/(\Delta - \delta_j) (\Delta - \delta_{j'})| \ll \sigma_{\delta}$, cross terms with $j \neq j'$ can be dropped, leaving us with:
\begin{equation}
\begin{aligned}
\kappa \mathcal{D} \left[\hat{a} - \sum_{j=1}^N \frac{g_j}{\Delta-\delta_j} \hat{s}_{j}^{-}\right]\rho &=\kappa \left(\hat{a} \rho \hat{a}^{\dagger}-\frac{1}{2}\hat{a}^{\dagger}\hat{a}\rho - \frac{1}{2}\rho\hat{a}^{\dagger}\hat{a}\right)\\
&+\kappa\sum_{j=1}^N \frac{|g_j|^2}{(\Delta - \delta_j)^2}\left(\hat{s}_{j}^{-}\rho \hat{s}_{j}^{+} - \frac{1}{2}\hat{s}_{j}^{+}\hat{s}_{j}^{-}\rho - \frac{1}{2}\rho\hat{s}_{j}^{+}\hat{s}_{j}^{-}\right)\\
&= \kappa \mathcal{D}[\hat{a}]\rho + \sum_{j=1}^N \frac{\kappa |g_j|^2}{(\Delta - \delta_j)^2}\mathcal{D}[\hat{s}_{j}^{-}]\rho.
\end{aligned}
\end{equation}
This is essentially the original single-photon loss, and a perturbative spin $T_1$ relaxation. The relaxation is \textit{single-spin} only -- the cancellation of the cross terms has suppressed the collective nature of the resonator-assisted loss, thus mitigating an $N$-dependent superradiant enhancement of the associated decay rate.

We can also look at the intrinsic loss of the spins, whose operators transform into,
\begin{equation}
\begin{aligned}
\hat{s}_{j}^{-} &\to \hat{s}_{j}^{-} + \frac{g_j^{*}}{\Delta-\delta_j}[\hat{a}\hat{s}_{j}^{+},\hat{s}_{j}^{-}]\\
&=\hat{s}_{j}^{-} + \frac{2g_j^{*}}{\Delta-\delta_j}\hat{a}\hat{s}_{j}^{z}.
\end{aligned}
\end{equation}
Similar to above, after going into the rotating-frame and dropping fast-rotating terms this approximately leads to intrinsic loss dissipators,
\begin{equation}
\gamma_{-}\sum_{j=1}^N \mathcal{D}[\hat{s}_{j}^{-}]\rho \to \gamma_{-}\sum_{j=1}^N \mathcal{D}[\hat{s}_{j}^{-}] \rho + \sum_{j=1}^N \frac{4\gamma_{-}|g_j|^2}{(\Delta-\delta_j)^2}\mathcal{D}[\hat{a}\hat{s}_{j}^{z}]\rho.
\end{equation}
For the purposes of dispersive measurement we can neglect the second term, since it conserves the polarization of each spin $\langle\hat{s}_j^{z}\rangle$, and doesn't strongly affect the resonator field because it's a weak correction to the raw resonator decay rate $\kappa$. The final dispersive model we get is thus
\begin{equation}
\begin{aligned}
\hat{H}_{\mathrm{SW}} &=\Delta \hat{a}^{\dagger}\hat{a} + \sum_{j=1}^N \delta_j \hat{s}_{j}^{z} -2\hat{a}^{\dagger}\hat{a} \sum_{j=1}^N \chi_j \hat{s}_{j}^{z},\\
\frac{d}{dt}\rho &= - i [\hat{H}_{\mathrm{SW}},\rho] + \kappa \mathcal{D}[\hat{a}]\rho + \sum_{j=1}^N \gamma_j \mathcal{D}[\hat{s}_{j}^{-}]\rho,\\
\end{aligned}
\end{equation}
where the coefficients are,
\begin{equation}
\chi_j = \frac{|g_j|^2}{\Delta -\delta_j},\>\>\>\>\gamma_j = \gamma_{-} + \frac{\kappa |g_j|^2}{(\Delta - \delta_j)^2}.
\end{equation}
This is the model reported in the main text.

As a final comment, our treatment assumes the spin detunings $\delta_j$ to be distributed over a wide enough range of frequencies $\delta_j$ (e.g. on the order of $\sim 1$ MHz linewidth) that any two individual spins are too far in frequency to have resonator-mediated interactions or decay. It is for this same reason that we assume the perturbation theory to be valid when the single-spin coupling is small compared to the detuning, $|g_j| \sqrt{\overline{n}} \ll |\Delta-\delta_j|$, without any collective enhancement proportional to the number of spins $N$. For a sufficiently large $N$, each spin may have a few neighbours close enough in frequency space that this assumption does not hold. The dominant effect of a few spins close in frequency is an enhancement of Purcell decay rate. Instead of a full $N$-fold amplification, having $2-3$ neighbours implies a $2-3$ times faster Purcell rate. For the parameter regimes we are interested in, such amplification will not change the overall outcome. Some amount of resonator-mediated flip-flop spin-spin interactions will also not affect our results, since the rates will be too slow to change the measurement outcome.

%%%%%
\section{Integrated homodyne quadrature}
\label{app_InputOutputHomodyne}
%%%%%

In this appendix, we derive our integrated homodyne measurement. We start with the dispersive model, and work in a frame rotating at the resonator frequency, thus taking $\hat{a} \to e^{i \Delta t}\hat{a}$. This removes the detuning Hamiltonian term $\Delta\hat{a}^{\dagger}\hat{a}$, but leaves all other Hamiltonian terms and dissipators unchanged since we are already in the dispersive regime. The resulting model reads,
\begin{equation}
\begin{aligned}
\hat{H}_{\mathrm{SW}} &=\sum_j \delta_j \hat{s}_{j}^{z} -2\hat{a}^{\dagger}\hat{a} \tilde{S}^{z},\\
\frac{d}{dt}\rho &= - i [\hat{H}_{\mathrm{SW}},\rho] + \kappa \mathcal{D}[\hat{a}]\rho + \sum_{j} \gamma_j \mathcal{D}[\hat{s}_{j}^{-}]\rho,
\end{aligned}
\end{equation}
where we have written
\begin{equation}
\tilde{S}^{z} = \sum_j \chi_j \hat{s}_j^z
\end{equation}
as the sum of the spin population operators weighted by their coupling to the resonator.

We next write the Langevin equation of motion for the resonator field under standard input-output theory, assuming the resonator is driven by an input field $\hat{a}_{\mathrm{in}}$ (e.g., a microwave signal driving the resonator), and that the resonator is overcoupled (i.e., the coupling to the driven mode $\kappa_c$ is approximately equal to total loss $\kappa$ from the resonator),
\begin{equation}
\label{eq_LangevinEquation}
\begin{aligned}
    \frac{d}{dt}\hat{a} &= -i [\hat{H}_{\mathrm{SW}},\hat{a}] + \kappa \left(\hat{a}^{\dagger}\hat{a} \hat{a} - \frac{1}{2}\{\hat{a}^{\dagger}\hat{a},\hat{a}\}\right) + \sum_j \gamma_j \left(\hat{s}_j^{+}\hat{a} \hat{s}_j^{-} - \frac{1}{2}\{\hat{s}_j^{+}\hat{s}_j^{-} , \hat{a}\}\right) - \sqrt{\kappa} \hat{a}_{\mathrm{in}},\\
    &=-\frac{\kappa}{2} \hat{a} + 2i \tilde{S}^{z}\hat{a} - \sqrt{\kappa}\hat{a}_{\mathrm{in}}.
\end{aligned}
\end{equation}
We assume the input field is a coherent drive resonant with the resonator frequency,
\begin{equation}
\hat{a}_{\mathrm{in}} = \alpha_{\mathrm{in}} + \hat{\xi},
\end{equation}
where $\alpha_{\mathrm{in}}$ the amplitude of the coherent state, and $\hat{\xi}$ a white-noise-distributed field characterizing its quantum fluctuations. We also assume that the resonator field is mostly in a coherent state of amplitude $\alpha$ itself, and expand about that coherent state via
\begin{equation}
\hat{a} = \alpha + \hat{b},
\end{equation}
where $\hat{b}$ characterizes quantum fluctuations. The equation of motion for the resonator fluctuations $\hat{b}$ reads
\begin{equation}
\frac{d}{dt}\hat{b} =  - \frac{\kappa}{2}\hat{b} + 2 i \tilde{S}^{z}\hat{b} + 2 i \alpha \tilde{S}^{z}-\sqrt{\kappa}\hat{\xi}+ \left(- \frac{\kappa}{2}\alpha - \sqrt{\kappa}\alpha_{\mathrm{in}}\right)
\end{equation}
Since $\hat{b}$ characterizes small fluctuations, and $\tilde{S}^{z}$ is also considered to be a small parameter as it contains the dispersive couplings $\chi_j$, we neglect the coupling term $2i \tilde{S}^z \hat{b}$ for simplicity (it can also be retained in the treatment, but will only provide a small correction). We also solve for the coherent field strength in the resonator by setting the last term in brackets to zero, which yields
\begin{equation}
\label{eq_unitsFixedLol}
\alpha = -\frac{2\alpha_{\mathrm{in}}}{\sqrt{\kappa}}.
\end{equation}
This assumes the resonator field follows the drive. The Langevin equation simplifies to
\begin{equation}
\frac{d}{dt}\hat{b} = - \frac{\kappa}{2}\hat{b} - \frac{4 i \alpha_{\mathrm{in}}}{\sqrt{\kappa}} \tilde{S}^{z} - \sqrt{\kappa}\hat{\xi}.
\end{equation}
This equation can be solved to yield
\begin{equation}
\hat{b}(t) = e^{- \frac{\kappa}{2}(t-t_0)}\hat{b}(t_0) +  e^{ - \frac{\kappa}{2}(t-t_0)}\int_{t_0}^{t}d\tau e^{\frac{\kappa}{2}(\tau-t_0)} \left[-\frac{4 i \alpha_{\mathrm{in}}}{\sqrt{\kappa}} \tilde{S}^{z}(\tau) - \sqrt{\kappa} \hat{\xi}(\tau)\right].
\end{equation}
We assume the dynamics of interest occurs at times much longer than the timescale of transient resonator dynamics $\sim 1 / \kappa$, meaning $t-t_0 \gg 1 /\kappa$ (i.e., we ignore the resonator ring-up time when the microwave drive is first turned on).
Hence, we take $t_0 \to - \infty$ and omit the first term involving the initial condition.

We also assume the resonator decay rate $\kappa$ is much faster than the spin dynamics timescales, thus $\tilde{S}^{z}(\tau)$ will change on a much slower timescale than $\sim 1 / \kappa$. The corresponding contribution to the integral is effectively weighted by a factor $\sim e^{-\frac{\kappa}{2}(t-\tau)}$, meaning that only contributions from $\tau \lesssim t$ on a timescale of $\sim 1 /\kappa$ will meaningfully have an effect. Since $\tilde{S}^{z}(\tau) $ varies slowly on this timescale, we assume it to be constant and replace it by $\tilde{S}^{z}(\tau) \approx \tilde{S}^{z}(t)$. The same approximation is made for the white noise of the drive, approximating $\hat{\xi(\tau)}\approx \hat{\xi}(t)$, which assumes that $\kappa \gg |\dot{\xi}(t)|$. Under these simplifying assumptions the integral can be evaluated explicitly,
\begin{equation}
\hat{b}(t) = -\frac{8 i \alpha_{\mathrm{in}}}{\kappa^{3/2}}\tilde{S}^{z}(t) - 2\frac{\hat{\xi}(t)}{\sqrt{\kappa}}.
\end{equation}
Under the input-output formalism, we can now obtain the outgoing field reflecting from the resonator (still assuming an overcoupled system $\kappa_c \approx \kappa$),
\begin{equation}
\begin{aligned}
\hat{a}_{\mathrm{out}}(t) &= \hat{a}_{\mathrm{in}}(t) + \sqrt{\kappa} \hat{a}(t),\\
&=- \frac{8 i \alpha_{\mathrm{in}}}{\kappa}\tilde{S}^{z}(t) -\alpha_{\mathrm{in}}- \hat{\xi}(t).
\end{aligned}
\end{equation}
We assume that the incoming field coherent amplitude is set by
\begin{equation}
\alpha_{\mathrm{in}} = i \sqrt{\dot{n}},
\end{equation}
where $\dot{n}$ is the average incoming photon flux ($\langle\hat{a}_{\mathrm{in}}^{\dagger}\hat{a}_{\mathrm{in}}^{\phantom{\dagger}}\rangle = \dot{n}$). The factor $i$ determines which quadrature of the outgoing resonator field we will need to measure via homodyne detection (i.e., $\hat{a}_{\mathrm{out}}+ \hat{a}_{\mathrm{out}}^{\dagger}$ versus $\hat{a}_{\mathrm{out}}- \hat{a}_{\mathrm{out}}^{\dagger}$).
We can also write this field strength in terms of the mean resonator photon number $\overline{n}$; the relation between the two can be inferred from Eq.~\eqref{eq_unitsFixedLol}, which yields
\begin{equation}
\overline{n} = \langle\hat{a}^{\dagger}\hat{a}\rangle = |\alpha|^2 = \frac{4 |\alpha_{\mathrm{in}}|^2}{\kappa} = \frac{4 \dot{n}} {\kappa}.
\end{equation}
The output field then reads
\begin{equation}
\hat{a}_{\mathrm{out}}(t) = \frac{4\sqrt{\overline{n}}}{\sqrt{\kappa}}\tilde{S}^{z}(t) - \frac{i}{2} \sqrt{\kappa\overline{n}}- \hat{\xi}(t).
\end{equation}
We measure a time-integrated homodyne current obtained by integrating the quadrature $\hat{I}_{\mathrm{out}}=\hat{a}_{\mathrm{out}} + \hat{a}_{\mathrm{out}}^{\dagger}$,
\begin{equation}
\begin{aligned}
\hat{M}(T) &= \int_0^{T} dt\> \hat{I}_{\mathrm{out}}(t)\\
&= \int_0^{T}dt\left[ \frac{8\sqrt{\overline{n}}}{\sqrt{\kappa}}\tilde{S}^{z}(t)  + \left(\hat{\xi}(t) + \hat{\xi}^{\dagger}(t)\right)\right].
\end{aligned}
\end{equation}
Taking expectation values and assuming $\hat{\xi}(t)$ to be white-noise distributed with $\langle\hat{\xi}(t)\rangle=0$, we get
\begin{equation}
\langle\hat{M}(T)\rangle =\frac{8\sqrt{\overline{n}}}{\sqrt{\kappa}} \int_0^{T}dt  \langle\tilde{S}^{z}(t)\rangle.
\end{equation}
This is the signal we seek to measure. 

The quantum noise about the signal is given by
\begin{equation}
\begin{aligned}
(\Delta M(T))^2 &=\langle\hat{M}(T)^2\rangle - \langle\hat{M}(T)\rangle^2\\
&= \left\langle \left( \int_0^T dt \left[ \frac{8 \sqrt{\overline{n}}}{\sqrt{\kappa}}\tilde{S}^{z}(t) + \left(\hat{\xi}(t) + \hat{\xi}^{\dagger}(t)\right)\right]\right)^2 \right\rangle -\left(\frac{8 \sqrt{\overline{n}}}{\sqrt{\kappa}}\right)^2\left[\int_0^{T}dt\langle \tilde{S}^{z}(t)\rangle\right]^2\\
&= \frac{64\overline{n}}{\kappa}  \left[ \int_0^T dt \int_0^T dt' \langle\tilde{S}^{z}(t) \tilde{S}^{z}(t')\rangle -\left(\int_0^{T}dt\langle \tilde{S}^{z}(t)\rangle\right)^2 \right] \\
&+ \frac{16 \sqrt{\overline{n}}}{\sqrt{\kappa}}\int_0^{T}dt \int_0^{T}dt' \langle\tilde{S}^{z}(t)\left[\hat{\xi}(t') + \hat{\xi}^{\dagger}(t')\right]\rangle\\
&+ \int_0^T dt \int_0^T dt' \left(\langle\hat{\xi}(t)\hat{\xi}(t')\rangle + \langle\hat{\xi}(t)\hat{\xi}^{\dagger}(t')\rangle + \langle\hat{\xi}^{\dagger}(t)\hat{\xi}(t')\rangle + \langle\hat{\xi}^{\dagger}(t)\hat{\xi}^{\dagger}(t')\rangle\right).
\end{aligned}
\end{equation}
Since $\hat{\xi}(t)$ is white-noise distributed, $\langle\hat{\xi}(t)\rangle=0$ and $\langle\hat{\xi}(t) \hat{\xi}^{\dagger}(t')\rangle = \delta(t-t')$ with all other correlators zero. This lets us drop the second line and evaluate the integral on the third line of the last equation, yielding
\begin{equation}
(\Delta M(T))^2 = \frac{64\overline{n}}{\kappa}  \left[ \int_0^T dt \int_0^T dt' \langle\tilde{S}^{z}(t) \tilde{S}^{z}(t')\rangle -\left(\int_0^{T}dt\langle \tilde{S}^{z}(t)\rangle\right)^2 \right] + T.
\end{equation}
We identify contributions from spin noise and shot noise:
\begin{equation}
(\Delta M(T))^2 = (\Delta M_{\mathrm{spin}}(T))^2 + (\Delta M_{\mathrm{shot}}(T))^2,
\end{equation}
each given by
\begin{equation}
\begin{aligned}
(\Delta M_{\mathrm{spin}}(T))^2 &= \frac{64\overline{n}}{\kappa}\sum_{j,j'=1}^N \chi_j \chi_{j'} \int_0^T dt \int_0^T dt' \left[\langle\hat{s}_j^{z}(t) \hat{s}_{j'}^{z}(t')\rangle - \langle\hat{s}_j^{z}(t)\rangle \langle\hat{s}_{j'}^{z}(t')\rangle \right],\\
(\Delta M_{\mathrm{shot}}(T))^2&= T,
\end{aligned}
\end{equation}
where we have reinserted $\tilde{S}^{z}(t) = \sum_{j}\chi_j(t)$.

%%%%%
\section{Spin dynamics and SNR under $T_1$ decay}
\label{app_SpinDynamics}
%%%%%

In this appendix we analytically solve the dynamics of spin population under $T_1$ decay (including two-time correlations), and use this to obtain analytic expressions for the SNR. The Heisenberg equations of motion for the spin population and correlator observables under the dispersive model are:
\begin{equation}
\begin{aligned}
\frac{d}{dt}\hat{s}_{j}^{z} &=  - \frac{\gamma_j}{2}(\mathbbm{1} + 2 \hat{s}_{j}^{z}),\\
\frac{d}{dt}\hat{s}_{j}^{z}\hat{s}_{j'}^{z} &= \begin{cases} - (\gamma_j + \gamma_{j'})\hat{s}_j^z \hat{s}_{j'}^z - \frac{\gamma_j}{2}\hat{s}_{j'}^{z} - \frac{\gamma_{j'}}{2}\hat{s}_{j}^{z} & j \neq j' \\ 0  &  j = j'\end{cases}
\end{aligned}
\end{equation}
These equations can be solved explicitly:
\begin{equation}
\begin{aligned}
\label{eq_onePoint}
\langle\hat{s}_{j}^{z}(t)\rangle &=  - \frac{1}{2}+\frac{e^{-\gamma_j t}}{2}\left(1 + 2\langle\hat{s}_{j}^{z}(0)\rangle\right)\\
\langle\hat{s}_{j}^{z}(t)\hat{s}_{j'}^{z}(t)\rangle &= \begin{cases}
e^{-(\gamma_j + \gamma_{j'})t}\langle\hat{s}_j^{z}\hat{s}_{j'}^{z}(0)\rangle - \frac{e^{-\gamma_{j} t}}{2}\left(1-e^{-\gamma_{j'}t}\right)\langle\hat{s}_{j}^{z}(0)\rangle - \frac{e^{-\gamma_{j'} t}}{2}\left(1-e^{-\gamma_{j}t}\right)\langle\hat{s}_{j'}^{z}(0)\rangle & j \neq j' \\
\quad\quad\quad\quad\quad\quad\quad\quad\quad\quad\quad\quad\quad\quad\quad\quad\quad\quad\quad\quad\quad+ \frac{1}{4}\left(1-e^{-\gamma_j t}\right)\left(1-e^{-\gamma_{j'}t}\right) &\\
\frac{1}{4} & j = j'.
\end{cases}
\end{aligned}
\end{equation}
To compute the variance we also need two-time correlators. For this we must invoke the quantum regression theorem, which provides a new differential equation for two-time two-point correlators from the one-point equation:
\begin{equation}
\label{eq_RegressionEquation}
\frac{d}{d\tau} \langle\hat{s}_j^{z}(t) \hat{s}_{j'}^{z}(t+\tau)\rangle = -\frac{\gamma_{j'}}{2}\left(\langle\hat{s}_j^{z}(t)\rangle + 2 \langle \hat{s}_j^{z}(t)\hat{s}_{j'}^{z}(t+\tau)\rangle\right),\>\>\>\tau \geq 0.
\end{equation}
We insert the first line of Eq.~\eqref{eq_onePoint} into Eq.~\eqref{eq_RegressionEquation}, use the second line of Eq.~\eqref{eq_onePoint} as the initial condition (for $\tau = 0$), solve the equation, then replace $\tau = t' - t$ (keeping in mind that we thus assume $t' \geq t$). This yields:
\begin{equation}
\label{eq_regressionSol}
\langle \hat{s}_j^{z}(t) \hat{s}_{j'}^{z}(t')\rangle = \begin{cases}
e^{-(\gamma_j t + \gamma_{j'}t')}\langle\hat{s}_j^{z}\hat{s}_{j'}^{z}(0)\rangle + \frac{e^{-\gamma_{j} t}}{2}\left(1-e^{-\gamma_{j'}t'}\right)\langle\hat{s}_{j}^{z}(0)\rangle + \frac{e^{-\gamma_{j'} t'}}{2}\left(1-e^{-\gamma_{j}t}\right)\langle\hat{s}_{j'}^{z}(0)\rangle & j \neq j' \\
\quad\quad\quad\quad\quad\quad\quad\quad\quad\quad\quad\quad\quad\quad\quad\quad\quad\quad\quad\quad\quad\quad+ \frac{1}{4}\left(1-e^{-\gamma_j t}\right)\left(1-e^{-\gamma_{j'}t'}\right) &\\
   \frac{1}{2}\left(\langle\hat{s}_j^{z}(0)\rangle + \frac{1}{2}\right)\left(e^{- \gamma_j t'} - e^{-\gamma_j t}\right) + \frac{1}{4} & j = j'.
\end{cases}
\end{equation}
We can also write the connected portion of the correlator:
\begin{equation}
\langle \hat{s}_j^{z}(t) \hat{s}_{j'}^{z}(t')\rangle - \langle\hat{s}_j^{z}(t)\rangle \langle\hat{s}_j^{z}(t')\rangle =\begin{cases}
e^{-(\gamma_j t + \gamma_{j'}t')}\left(\langle\hat{s}_j^{z}\hat{s}_{j'}^{z}(0)\rangle - \langle \hat{s}_j^{z}(0)\rangle \langle\hat{s}_{j'}^{z}(0)\rangle\right) & j \neq j'\\
-\left(\langle\hat{s}_j^{z}(0)\rangle + \frac{1}{2}\right)^2 e^{- \gamma_j (t+t')} + \left(\langle\hat{s}_j^{z}(0)\rangle + \frac{1}{2}\right) e^{-\gamma_j t'} & j = j'.
\end{cases}
\end{equation}
Unsurprisingly, the connected correlator vanishes for $j \neq j'$ if we start in a separable state $\langle\hat{s}_j^{z}\hat{s}_{j'}^z(0)\rangle = \langle\hat{s}_j^{z}(0)\rangle \langle\hat{s}_{j'}^{z}(0)\rangle$.

We can now integrate the observables. The one-point observable is:
\begin{equation}
\label{eq_integratedSpin1PObs}
\int_0^{T}dt \langle\hat{s}_j^{z}(t)\rangle =\frac{1}{\gamma_{j}}\left(\langle\hat{s}_j^{z}(0)\rangle + \frac{1}{2}\right)\left(1-e^{-\gamma_j T}\right) - \frac{T}{2}.
\end{equation}
For the two-point observable, we are integrating over both $t$ and $t'$ equally, but the two-point correlator requires $t' \geq t$. To keep our expressions valid, for the region of integration with $t' < t$ we must make the exchange $t \leftrightarrow t'$ and $j \leftrightarrow j'$ in the integrand. Performing the integration yields
\begin{equation}
\begin{aligned}
\label{eq_integratedSpin2PObs}
&\int_0^{T}dt \int_0^{T}dt' \left[ \langle\hat{s}_j^{z}(t) \hat{s}_{j'}^{z}(t')\rangle -\langle\hat{s}_j^{z}(t)\rangle \langle \hat{s}_{j'}^{z}(t')\rangle\right] \\
&=\begin{cases}
\frac{1}{\gamma_j \gamma_{j'}} \left(1-e^{-\gamma_j T}\right)\left(1-e^{-\gamma_{j'}T}\right) \left(\langle \hat{s}_j^{z}\hat{s}_{j'}^{z}(0)\rangle - \langle\hat{s}_{j}^{z}(0)\rangle \langle\hat{s}_{j'}^{z}(0)\rangle\right) & j \neq j' \\
\frac{1}{\gamma_j^2}\left(\langle \hat{s}_j^{z}(0)\rangle + \frac{1}{2}\right)\bigg[2\left(1-e^{-\gamma_j T}\right)- 2 e^{-\gamma_j T} \gamma_j T-\left(\langle \hat{s}_j^{z}(0)\rangle + \frac{1}{2}\right)\left(1-e^{-\gamma_j T}\right)^2 \bigg] & j = j'.
\end{cases}
\end{aligned}
\end{equation}

For a separable initial state
$\langle\hat{s}_j^{z}\hat{s}_{j'}^z(0)\rangle = \langle\hat{s}_j^{z}(0)\rangle \langle\hat{s}_{j'}^{z}(0)\rangle$, only contributions to the variance from $j = j'$ are non-zero. Under this assumption, we can explicitly write the spin noise contribution to the SNR of our measurement:
\begin{equation}
\label{eq_integratedSpin2PObsSeparable}
\begin{aligned}
(\Delta M_{\mathrm{spin}}(T))^2 &= \frac{64 \overline{n}}{\kappa}\sum_{j} \chi_j^2 \int_0^{T}dt \int_0^{T}dt' \left[ \langle\hat{s}_j^{z}(t) \hat{s}_{j}^{z}(t')\rangle -\langle\hat{s}_j^{z}(t)\rangle^2\right],\quad\quad \text{(separable initial state)}\\
&= \frac{64 \overline{n}}{\kappa}\sum_{j}\frac{\chi_j^2}{\gamma_j^2}\left(\langle\hat{s}_j^z(0)\rangle + \frac{1}{2}\right)\bigg[2\left(1-e^{-\gamma_j T}\right)-\left(\langle \hat{s}_j^{z}(0)\rangle + \frac{1}{2}\right)\left(1-e^{-\gamma_j T}\right)^2 - 2 e^{-\gamma_j T} \gamma_j T\bigg].
\end{aligned}
\end{equation}

%%%%%%%%%%%%%%%%
\section{Phase noise}
\label{app_PhaseNoise}
%%%%%%%%%%%%%%%%

Here we analyze the impact of phase noise on our protocol, either in the resonator itself or in driving the resonator with a microwave signal generator. Suppose that our resonator lowering operator has a phase that fluctuates in time,
\begin{equation}
\hat{a} \to \hat{a} e^{i \phi(t)},
\end{equation}
where $\phi(t)$ is a random phase with zero-mean and Gaussian temporal correlations~\cite{diosi2008laser, rabl2009phase}. For simplicity, we consider a noise process with a Lorentzian noise spectrum, for which classical averages of this phase factor can be written explicitly as:
\begin{equation}
\label{eq_ClassicalAvg}
\begin{aligned}
\langle \dot{\phi}(t) \dot{\phi}(t') \rangle_{\mathrm{classical}} &= \Gamma_L \gamma_c e^{- \gamma_c |t-t'|},\\
\langle \dot{\phi}(t) \rangle_{\mathrm{classical}}  &= 0,
\end{aligned}
\end{equation}
where $\Gamma_L$ is the Lorentzian width of the resonator phase noise (i.e., the characteristic frequency width of its phase noise spectrum), and $\gamma_c$ the noise correlation time. In the white noise limit $\gamma_c \to \infty$, we have $\langle \dot{\phi}(t) \dot{\phi}(t') \rangle_{\mathrm{classical}} \approx 2 \Gamma_L \delta(t-t')$.

We can go into a frame rotating at this new adjusted frequency by making a unitary transformation,
\begin{equation}
\hat{U}(t) = e^{i \phi(t) \hat{a}^{\dagger}\hat{a}},
\end{equation}
which gets rid of the phase dependence in the resonator annihilation operator. The usual price we pay is an extra term in the Hamiltonian,
\begin{equation}
-i\hat{U}^{\dagger}(t) \frac{d}{dt}\hat{U}(t) = \dot{\phi}(t) \hat{a}^{\dagger}\hat{a},
\end{equation}
i.e., a fluctuating resonator frequency.

The Langevin equation for our measurement scheme (from Eq.~\eqref{eq_LangevinEquation}) with the addition of this new Hamiltonian term becomes,
\begin{equation}
    \frac{d}{dt}\hat{a}=\left[-\frac{\kappa}{2} - i \dot{\phi}(t)\right]\hat{a} + 2i \tilde{S}^{z}\hat{a} - \sqrt{\kappa}\hat{a}_{\mathrm{in}}.
\end{equation}
We proceed to solve this equation in the same way as the previous sections: expand both the incoming field $\hat{a}_{\mathrm{in}} = \alpha_{\mathrm{in}} + \hat{\xi}$ and the resonator field $\hat{a} = \alpha + \hat{b}$, ignore resonator ring-up time to omit 
contributions from initial conditions, and assume the resonator and drive coherent amplitudes are locked via,
\begin{equation}
\alpha = - \frac{\sqrt{\kappa}\alpha_{\mathrm{in}}}{\frac{\kappa}{2} + i \dot{\phi}(t)},
\end{equation}
in analogy to Eq.~\eqref{eq_unitsFixedLol} but containing the extra contribution from the phase-noise term. This assumption implies that the timescale for phase noise fluctuations is much slower than the resonator decay rate,
\begin{equation}
\kappa \gg |\dot{\phi}(t)|.
\end{equation}
The formal solution for the resonator fluctuations $\hat{b}(t)$ then reads,
\begin{equation}
\hat{b}(t) =\int_{-\infty}^{t} d\tau e^{-\frac{\kappa}{2}(t - \tau)- i [\phi(t) - \phi(\tau)]} \left[-\frac{2i \sqrt{\kappa}\alpha_{\mathrm{in}}}{\frac{\kappa}{2} + i \dot{\phi}(\tau)}\tilde{S}^{z}(\tau) - \sqrt{\kappa}\hat{\xi}(\tau)\right].
\end{equation}

Still working under the assumption
$\kappa \gg |\dot{\phi}(t)|$, we can approximate $\phi(\tau) \approx \phi(t)$ in the integrals. As before, we may do the same for the spin dynamics $\tilde{S}^z(\tau) \approx \tilde{S}^z (t)$ and the white noise of the drive $\hat{\xi}(\tau) \approx \hat{\xi}(t)$. The only $\tau$ dependence is thus in the exponential, and hence we can do the integral analytically,
\begin{equation}
\hat{b}(t) = \frac{2}{\kappa}\left[-\frac{2i \sqrt{\kappa}\alpha_{\mathrm{in}}}{\frac{\kappa}{2} + i \dot{\phi}(t)}\tilde{S}^{z}(t) - \sqrt{\kappa}\hat{\xi}(t)\right].
\end{equation}
Under the input-output formalism, as before we calculate the outgoing field,
\begin{equation}
\begin{aligned}
\hat{a}_{\mathrm{out}}(t) &= \hat{a}_{\mathrm{in}}(t) + \sqrt{\kappa} \hat{a}(t)\\
&=-\frac{1-\frac{2 i \dot{\phi}(t)}{\kappa}}{1+\frac{2 i \dot{\phi}(t)}{\kappa}} \alpha_{\mathrm{in}} - \frac{4i \alpha_{\mathrm{in}}}{\frac{\kappa}{2} + i \dot{\phi}(t)}\tilde{S}^{z}(t) - \hat{\xi}(t),\\
\end{aligned}
\end{equation}
We next write the incoming field coherent amplitude in terms of the resonator photon number $\overline{n}$,
\begin{equation}
\alpha_{\mathrm{in}} = \frac{i \sqrt{\kappa \overline{n}}}{2},
\end{equation}
with which the output field becomes:
\begin{equation}
\hat{a}_{\mathrm{out}}(t) =-\frac{i}{2}\frac{1-\frac{2 i \dot{\phi}(t)}{\kappa}}{1+\frac{2 i \dot{\phi}(t)}{\kappa}} \sqrt{\kappa \overline{n}}+ \frac{2\sqrt{\kappa \overline{n}}}{\frac{\kappa}{2} + i \dot{\phi}(t)}\tilde{S}^{z}(t) - \hat{\xi}(t).
\end{equation}
The integrated homodyne quadrature is
\begin{equation}
\begin{aligned}
\hat{M}(T) &=\int_0^{T} dt \left(\hat{a}_{\mathrm{out}}(t) + \hat{a}_{\mathrm{out}}^{\dagger}(t) \right)\\
&=\int_0^{T}dt\left[\frac{4\kappa^{3/2}\sqrt{\overline{n}}}{\kappa^2 + 4 \dot{\phi}^2(t)}\left(2\tilde{S}^{z}(t) - \dot{\phi}(t)\right) + \left(\hat{\xi}(t) + \hat{\xi}^{\dagger}(t)\right)\right],
\end{aligned}
\end{equation}
whose expectation value is the same as the outcome without phase noise,
\begin{equation}
\begin{aligned}
\langle\hat{M}(T)\rangle = \frac{8 \sqrt{\overline{n}}}{\sqrt{\kappa}} \int_0^{T}dt\langle \tilde{S}^{z}(t)\rangle,
\end{aligned}
\end{equation}
under the assumption that the phase noise has zero mean derivative $\dot{\phi}(t)=0$, as do the quantum fluctuations in the shot noise $\langle\hat{\xi}(t)\rangle=0$. We also dropped the $\dot{\phi}^2(t)$ in the denominator since we still assume a fast resonator loss $\kappa \gg |\dot{\phi}(t)|$.

The quantum noise about the signal is given by
\begin{equation}
\begin{aligned}
(\Delta M(T))^2  &= \langle\hat{M}^2(T)\rangle - \langle \hat{M}(T)\rangle^2\\
&=\frac{16 \overline{n}}{\kappa}\int_0^T dt \int_0^T dt' \bigg[4\langle\tilde{S}^{z}(t) \tilde{S}^{z}(t')\rangle - 4\langle\tilde{S}^{z}(t)\rangle \dot{\phi}(t') +  \dot{\phi}(t)\dot{\phi}(t')\bigg]+ T\\
&- \frac{16 \overline{n}}{\kappa} \left[\int_0^T dt \left(-2 \langle\tilde{S}^{z}(t)\rangle 
 + \dot{\phi}(t)\right)\right]^2 .
\end{aligned}
\end{equation}
In writing this expression, we again dropped all cross terms involving $\langle\hat{\xi}(t)\rangle=0$, dropped terms involving $\dot{\phi}^2(t)$ in the denominators, and used the two-point correlator $\langle\hat{\xi}(t) \hat{\xi}^{\dagger}(t')\rangle = \delta(t-t')$ with all other correlators zero since $\hat{\xi}(t)$ is white noise distributed.
To evaluate the integrals involving the phase noise contributions, we can now perform a classical average via Eqs.~\eqref{eq_ClassicalAvg} in the white-noise limit, namely $\langle\dot{\phi}(t)\rangle_{\mathrm{classical}}=0$, $\langle\dot{\phi}(t)\dot{\phi}(t')\rangle_{\mathrm{classical}}=2 \Gamma_L \delta(t-t')$, which yields:
\begin{equation}
(\Delta M(T))^2 =\frac{64 \overline{n}}{\kappa} \int_0^T dt \int_0^T dt' \left[\langle\tilde{S}^{z}(t) \tilde{S}^{z}(t')\rangle - \langle\tilde{S}^{z}(t) \rangle \langle\tilde{S}^{z}(t')\rangle\right]+ \left(1 +  \frac{32\overline{n} \Gamma_L}{\kappa}\right) T,\\
\end{equation}
We find that the leading-order effect of the phase noise is to thus \textit{increase} the shot noise by $T \to (1+32 \overline{n} \Gamma_L / \kappa)T$. The measurement quality parameter $\lambda$ thus inherits this factor, and becomes,
\begin{equation}
\lambda \to \frac{\lambda}{1+\frac{32 \overline{n}\Gamma_L}{\kappa}}.
\end{equation}

%%%%%%%%%%%%%%%%
\section{Finite number of experimental runs}
\label{app_FiniteRuns}
%%%%%%%%%%%%%%%%

Here we explore the effects of a finite number $N_{\mathrm{runs}}$ of experimental runs. If the goal is simply to show that a measurement is spin-projection-noise limited, the first task is to demonstrate that $(\Delta M_{\mathrm{spin}}(T))^2 > (\Delta M_{\mathrm{shot}}(T))^2$.
The total measurement variance $(\Delta M(T))^2$ discussed thus far is a total variance assuming an infinite number of experimental runs. A real experiment with data from a finite number $N_{\mathrm{runs}}$ of experimental runs will instead have a sample variance $(\Delta M_{\mathrm{sample}}(T))^2$ that may differ from the true $(\Delta M(T))^2$. For a Gaussian distributed random variable, the average difference between the true and sample variance will scale as $\mathbbm{E}[|(\Delta M(T))^2-(\Delta M_{\mathrm{sample}}(T))^2|]\sim (\Delta M(T))^2/\sqrt{N_{\mathrm{runs}}}$. To distinguish the presence of additional noise caused by the spins above the shot noise $(\Delta M_{\mathrm{spin}}(T))^2= (\Delta M(T))^2 - (\Delta M_{\mathrm{shot}}(T))^2$, thus signifying a measurement is spin projection resolved, the difference between sample and true variance must be smaller than this additional noise. Mathematically, to see the spin noise, we must have:
\begin{equation}
\frac{(\Delta M(T))^2}{\sqrt{N_{\mathrm{runs}}}} \leq (\Delta M_{\mathrm{spin}}(T))^2.
\end{equation}
For homogeneous parameters, this condition simplifies to:
\begin{align}
N_{\mathrm{runs}} \geq \left(1 + \frac{1}{\lambda }\frac{\gamma T}{\left(1-e^{-\gamma T}\right)\left(3+e^{-\gamma T}\right)-4e^{-\gamma T}\gamma T}\right)^2.
\label{eq:Nruns}
\end{align}

At short times $\gamma T \ll 1$ and  $\lambda \gg 1$, we can expand this expression to find:
\begin{equation}
N_{\mathrm{runs}}\geq 1+\frac{1}{\lambda^2\gamma^2 T^2}+\frac{2}{\lambda\gamma T}+\mathcal{O}(\gamma T,\frac{1}{\lambda}),
\end{equation}
where all terms proportional to $1/\lambda$ or $\gamma T$ (but not the ratio of the two) are dropped. We observe from this power series that there is a natural competition between the dimensionless $\lambda$ and the evolution time $\gamma T$ under $T_1$ decay.

We may also numerically optimize the full expression, Eq.~\eqref{eq:Nruns}, to find that the optimal collection time is $\gamma T \approx 2.1$, independent of $\lambda$. The number of runs at this optimal time is $N_{\mathrm{runs}}\approx (1+\frac{1.2}{\lambda})^2$.

Since each experimental run involves a collection time $T$, if the time needed to reset the experiment is short compared to $T$, we may instead seek to minimize the product $N_{\mathrm{runs}}T$ (i.e. the overall experimental time over all shots). In this case the optimal collection time now depends on $\lambda$. For $\lambda \gg 1$, the optimal time approximately satisfies the simple relation $\gamma T \approx \frac{1}{\lambda}$; the measurement quality parameter directly sets how long to measure. However, real experiments may have other constraints that limit $T$ to shorter timescales -- in this case, the general expression for $N_{\mathrm{runs}}$ may be used.

%%%%%%%%%%%%%%%%
\section{Entangled initial state}
\label{app_Entanglement}
%%%%%%%%%%%%%%%%
%%%%%
\subsection{Difference in variance due to entanglement}
%%%%%

In this Appendix we consider the performance of the measurement protocol for a non-separable entangled initial state, such as a spin-squeezed state. In particular, suppose we start with all spins in $+x$, and perform an entangling unitary with a one-axis twisting (OAT) Hamiltonian for an entangling time $t_{\mathrm{sqz}}$:
\begin{equation}
\begin{aligned}
\ket{\psi_{\mathrm{sqz}}} &= e^{-i t_{\mathrm{sqz}}\hat{H}_{\mathrm{OAT}}} \prod_j \frac{1}{\sqrt{2}}\left(\ket{\uparrow}_j + \ket{\downarrow}_j\right),\\
\hat{H}_{\mathrm{OAT}} &= \frac{1}{N}\sum_{j,j'}\hat{s}_j^{z}\hat{s}_{j'}^{z}.
\end{aligned}
\end{equation}
The entangling time $t_{\mathrm{sqz}}$ is assumed to absorb the energy unit of the interaction Hamiltonian prefactor for simplicity.
We also normalize the interaction by $1/N$ so it is extensive in energy. The maximal and minimal variances $V_{+}$ and $V_{-}$ of the resulting spin phase space distribution are given by~\cite{kitagawa1993seminal}
\begin{equation}
\begin{aligned}
V_{\pm}&=\frac{N}{4}+\frac{N}{16}\left(N-1\right)\left[1-\cos^{N-2}\left(\frac{2t_{\mathrm{sqz}}}{N}\right)\pm\sqrt{\left(1-\cos^{N-2}\left(\frac{2t_{\mathrm{sqz}}}{N}\right)\right)^2+16\sin^2\left(\frac{t_{\mathrm{sqz}}}{N}\right)\cos^{2N-4}\left(\frac{t_{\mathrm{sqz}}}{N}\right)}\right]\\
&=\frac{N}{4}+\frac{N}{8}\left(t_{\mathrm{sqz}}^2\pm t_{\mathrm{sqz}}\sqrt{4+t_{\mathrm{sqz}}^2}\right) +\mathcal{O}\left(t_{\mathrm{sqz}}^2,\frac{1}{N}\right).
\end{aligned}
\end{equation}
Hereafter, for simplicity, we will assume that the amount of squeezing generated is small, i.e., we assume that the entangling time does not scale with $N$. For OAT, optimal squeezing occurs at $t_{\mathrm{sqz}}\sim N^{1/3}$; we will consider $t_{\mathrm{sqz}}$ much smaller than this. After the entangling step, the spins are rotated by another unitary $\ket{\psi_{\mathrm{sqz},\theta_{\mathrm{sqz}}}} = e^{-i \theta_{\mathrm{sqz}}\hat{S}^{x}}\ket{\psi_{\mathrm{sqz}}}$ with an angle $\theta_{\mathrm{sqz}}$ such that $\bra{\psi_{\mathrm{sqz},\theta_{\mathrm{sqz}}}}\hat{S}^{z}\hat{S}^{z}\ket{\psi_{\mathrm{sqz},\theta_{\mathrm{sqz}}}} = V_{-}$ (if we study a squeezed state), or $\bra{\psi_{\mathrm{sqz},\theta_{\mathrm{sqz}}}}\hat{S}^{z}\hat{S}^{z}\ket{\psi_{\mathrm{sqz},\theta_{\mathrm{sqz}}}}=V_{+}$ (for an anti-squeezed state). The spin contrast of the state is given by
\begin{equation}
\bra{\psi_{\mathrm{sqz},\theta_{\mathrm{sqz}}}}\hat{S}^{x}\ket{\psi_{\mathrm{sqz},\theta_{\mathrm{sqz}}}} = \frac{N}{2}\cos^{N-1}\left(\frac{t_{\mathrm{sqz}}}{N}\right) = \frac{N}{2} + \mathcal{O}\left(t_{\mathrm{sqz}}^2\right).
\end{equation}
For weak squeezing, the Wineland squeezing parameter is then just proportional to the variance,
\begin{equation}
\xi^2 = N\frac{\bra{\psi_{\mathrm{sqz},\theta_{\mathrm{sqz}}}}\hat{S}^{z}\hat{S}^{z}\ket{\psi_{\mathrm{sqz},\theta_{\mathrm{sqz}}}}}{\bra{\psi_{\mathrm{sqz},\theta_{\mathrm{sqz}}}}\hat{S}^{x}\ket{\psi_{\mathrm{sqz},\theta_{\mathrm{sqz}}}}} \approx \frac{4\bra{\psi_{\mathrm{sqz},\theta_{\mathrm{sqz}}}}\hat{S}^{z}\hat{S}^{z}\ket{\psi_{\mathrm{sqz},\theta_{\mathrm{sqz}}}}}{N}.
\end{equation}

The spin contributions to the noise of our measurement protocol contain an extra term [$j\neq j'$ term on last line of Eq.~\eqref{eq_integratedSpin2PObs}] that encodes a non-separable initial state. For homogeneous system parameters and the squeezed state described above, we can write this extra contribution as,
\begin{equation}
\delta (\Delta M_{\mathrm{sqz}}(T))^2= (\xi^2-1)\frac{16 N\overline{n}\chi^2}{\kappa\gamma^2}\left(1-e^{-\gamma T}\right)^2.
\end{equation}
The overall variance of the measurement with the entangled state is then (assuming $N \gg 1$)
\begin{equation}
(\Delta M_{\mathrm{sqz}}(T))^2= \frac{64 N\overline{n}\chi^2}{\kappa\gamma^2}\left[1-e^{-\gamma T} - e^{-\gamma T}\gamma T - \frac{1}{4}\left(1-e^{-\gamma T}\right)^2\right]+ T + \delta (\Delta M_{\mathrm{sqz}}(T))^2.
\end{equation}

%%%
\subsection{Finite number of experimental runs}
%%%

We now again consider an experiment with a finite number of runs $N_{\mathrm{runs}}$, which only has a sample variance for each collection time $T$ rather than the true variance. We can again write an analytic condition for the number of experimental runs needed to resolve the difference in variance $\delta (\Delta M_{\mathrm{sqz}}(T))^2$ due to entanglement. This time we require the average difference in true and sample variance, $\mathbbm{E}[|(\Delta M_{\mathrm{sqz}}(T))^2 -(\Delta M_{\mathrm{sqz,sample}}(T))^2|] \sim (\Delta M_{\mathrm{sqz}}(T))^2 / \sqrt{N_{\mathrm{runs}}}$, to be smaller than the change in variance due to entanglement $\delta (\Delta M_{\mathrm{sqz}}(T))^2$:
\begin{equation}
\frac{(\Delta M_{\mathrm{sqz}}(T))^2}{\sqrt{N_{\mathrm{runs}}}} \leq \delta (\Delta M_{\mathrm{sqz}}(T))^2.
\end{equation}
This condition yields a minimum number of experimental runs needed to see the effects of entanglement:
\begin{equation}
\label{eq_MinShots}
N_{\mathrm{runs}}\geq \left(1+\frac{4\left[1-e^{-\gamma T}-e^{-\gamma T}\gamma T-\frac{1}{4}\left(1-e^{-\gamma T}\right)^2\right]+\frac{1}{\lambda} \gamma T}{\left(\xi^2-1\right)\left(1-e^{-\gamma T}\right)^2}\right)^2.
\end{equation}
In the limit $\xi^2 \to \infty$, the requirement reduces to $N_{\mathrm{runs}}\geq 1$ as an infinitely large increase in variance can be resolved with a single run. The condition can also be expanded at short times to yield,
\begin{equation}
N_{\mathrm{runs}} \geq \left(1+\frac{\frac{1}{\lambda\gamma T}+\frac{1}{\lambda}+1+\frac{8+\frac{5}{\lambda}}{12}\gamma T}{\xi^2-1}\right)^2 +\mathcal{O}
\left(\gamma^2 T^2\right).
\end{equation}
The above function has a minimum at:
\begin{equation}
\gamma T_{\mathrm{min}} = \frac{2\sqrt{3}}{\sqrt{8\lambda+5}},
\end{equation}
which indicates an optimal time at which the number of needed experimental runs is minimized. If we instead seek to minimize total time $N_{\mathrm{runs}}T$, then the optimum time would instead be approximately,
\begin{equation}
    \gamma T_{\mathrm{min,total}} = \frac{1}{1+\lambda\xi^2}.
\end{equation}
As before, however, an experiment may face practical timescale constraints restricting the integration time to $\gamma T \ll 1$, in which case one may use Eq.~\eqref{eq_MinShots} to determine the minimum amount of experimental shots that will be needed.

\end{document}